\documentclass[journal,10pt]{IEEEtran}
\usepackage{graphicx,amsmath,amssymb}
\usepackage{booktabs}
\usepackage{threeparttable}

\usepackage{array}    
\usepackage{makecell} 
\usepackage{booktabs} 

\usepackage{array}
\usepackage{cite}
\usepackage{subfigure}
\usepackage{citesort}
\usepackage{fancyhdr}
\usepackage{mdwmath}
\usepackage{mdwtab}
\usepackage{balance}
\usepackage{caption2}
\usepackage{xcolor}
\usepackage{multirow}
\usepackage{bm}
\usepackage[T1]{fontenc}
\usepackage{enumerate}
\usepackage{algorithm}
\usepackage{amsmath}
\usepackage{algorithmic}
\usepackage{flafter}
\usepackage{epstopdf}
\usepackage{url}
\newtheorem{remark}{\textbf{Remark}}
\newtheorem{observation}{\textbf{Observation}}
\newtheorem{proposition}{\textbf{Proposition}\bfseries}
\newtheorem{theorem}{Theorem}

\newtheorem{lemma}{Lemma}

\newtheorem{corollary}{\textbf{Corollary}\bfseries}

\newtheorem{assumption}{Assumption}

\newenvironment{proof_black}{{ \it \textbf{Proof}:}}{$\hfill \blacksquare$\par}
\begin{document}
\title{Tag-based Physical-Layer Authentication Against Message Interference}

\author{Lei Yao,~
        Boxiang He,~
        Shilian Wang,~
        Enyu Shi,~
        and Chau Yuen,~\IEEEmembership{Fellow,~IEEE}
        
\thanks{L. Yao, B. He, and S. Wang are with the College of Electronic Science and
	Technology, National University of Defense Technology, Changsha 410003, P. R. China. (email: yaolei11103@163.com; boxianghe1@bjtu.edu.cn; wangsl@nudt.edu.cn).}
\thanks{E. Shi is with the State Key Laboratory of Advanced Rail Autonomous Operation, and also with the School of Electronics and Information Engineering, Beijing Jiaotong University, Beijing 100044, P. R. China. (e-mail: enyushi@bjtu.edu.cn).}
\thanks{C. Yuen is with the School of Electrical and Electronics Engineering, Nanyang Technological University, Singapore 639798 (e-mail:  chau.yuen@ntu.edu.sg).}
}
\maketitle
\begin{abstract}
 Tag-based Physical-Layer Authentication (PLA) has attracted significant attention in recent years due to its low complexity, high security, and low latency. Traditional tag-based PLA schemes typically estimate tags by decoding the message and then subtracting the estimation of the message from the received signal. However, these approaches suffer from two main limitations. First, decoding errors introduce message interference that degrades authentication performance. Second, the analytical complexity of decoding errors leads to sub-optimal threshold settings, thereby limiting detection probability. To address these limitations, this paper proposes a Tag-Based Challenge-Response (TBCR) scheme and a Series Cancellation Authentication (SCA) scheme. Specifically, in the TBCR scheme, the tags are superimposed on a forwarded challenge signal, enabling the receiver to estimate tags by removing the known challenge signal rather than relying on decoding. However, the challenge-response mechanism introduces extra noise. Here, we propose the SCA scheme without the noise interference, where both the series signal generation and cancellation modules are well-designed to generate authentication signals and estimate tags, respectively. Furthermore, we derive the closed-form expressions to evaluate the robustness and security of both proposed schemes. Notably, on one hand, the optimal threshold and detection probability are derived, which theoretically reveal that the SCA scheme always achieves the ideal detection performance, while the TBCR scheme does so in the absence of noise at Alice. On the other hand, the TBCR scheme provides enhanced security at high Signal-to-Noise Ratio (SNR) regions with fewer keys. Theoretical analysis and simulation demonstrate that both proposed schemes significantly outperform the benchmarks in detection probability with reduced time complexity. 
\end{abstract}


\section{Introduction}

\IEEEPARstart {D}{ue} to the proliferation of various mobile devices and the development of the Internet-of-Things (IoT), wireless communication has become an indispensable part of modern life. It is projected that the total number of global mobile broadband users will reach 17.1 billion with the traffic demands per user also continuously increasing by 2030 \cite{6G}. The rapid advancement of wireless communication facilitates the convenience and efficiency of information transmission. However, it simultaneously introduces challenges in data security and network authentication \cite{lhl,Wu,YanChao}. On one hand, traditional authentication relies on cryptographic methods at the upper layers for identity verification using passwords or tokens. However, corresponding authentication mechanisms are absent at the physical layer. Thus, adversaries can easily launch attacks on massive data and users through the physical medium because of the openness of wireless communication \cite{MITM,spoof}. On the other hand, the security of the Upper-Layer Authentication (ULA) guaranteed by cryptography has significantly diminished with the development of computing power \cite{Quan1}. Thus, the Physical-Layer Authentication (PLA) is garnering widespread attention as a supplement to the ULA\cite{XieSurvey,PLAsurvey1,PLAsurvey2}.

Distinct from the ULA, PLA leverages the physical characteristics of wireless signals to verify the identities of communication parties and it can be categorized into two types: passive and active schemes \cite{XieSurvey,HeSEII}. Passive schemes compare and identify physical-layer characteristics for authentication. Specifically, passive PLA schemes are usually based on hardware impairments (say, frequency offset \cite{CFO1}, clock skew \cite{TimeSkew1,TimeSkew2}, and I/Q imbalance \cite{IQ1,IQ2}), or on channel characteristics (say, channel phase responses \cite{ChannelRes1,ChannelRes2} and received signal strength \cite{RSS}). The hardware-impairment-based passive schemes face the issue of indistinguishable device features because of the advancements in electronic component manufacturing processes, and precise measurements of channel information are required for channel-characteristic-based passive schemes. Therefore, hardware requirements for the passive scheme are relatively higher \cite{HardWareIm}. In contrast, most of the active PLA schemes utilize the secret keys to superimpose a tag on the message or insert a tag into the message for authentication, known as the tag-based PLA schemes. Tag-based PLA schemes are more adaptable and can interfere with the adversary by flexibly adjusting the signal parameters, such as the power and the position of the tag.

There has been an increasing amount of research on the tag-based PLA schemes in recent years. The tag-based PLA schemes employ decoding methods at the authenticator (Bob) to recover the message, and subsequently estimate the tag by removing the estimated message from the received signal. Paul et al. propose a fundamental tag-based PLA scheme known as the SUPerimposition (SUP) scheme \cite{PLA2008}, which generates the authentication signal by carefully superimposing a tag on the modulated signal. Kumar et al. leverage the redundancy introduced by Precoded Duobinary Signaling (P-DS) to embed an authentication signal into the message, proposing an authentication scheme called the P-DS for authentication \cite{PDSA}. Xie et al. propose a generic PLA security model and design a new evaluation metric called the Probability of Security Authentication (PSA) \cite{XiePSA}. The PSA compares the reliable authentication performance between Bob and the attacker (Eve), providing a holistic assessment of the robustness, compatibility, and security. To address the additional overhead and vulnerability introduced by broadcasting the parameters of the tag-based PLA scheme, a Blind Tag-based  PLA (BTP) scheme is proposed in \cite{BTP} and \cite{XieBlind},  where Bob does not need to extract the tag from the received signal, thereby reducing the complexity at the receiver. A novel slope authentication scheme is proposed by constructing the transmitted signal \cite{XieSlope}. Specifically, the tag controls the signal grouping and the power allocation, and Bob can offset the influence of channel fading by constructing differential test statistics. 

Additionally, the Challenge-Response (CR) mechanism has been applied to PLA. In the CR mechanism, the authenticated party (Alice) utilizes a secret key to transform Bob's challenge signal into a response signal. Thus, Bob can authenticate Alice with the shared secret key and the known challenge signal. Shan et al. propose a physical-layer Challenge-Response Authentication Mechanism (CRAM)\cite{CRAM}, where Alice generates a proportional response signal to mitigate the effects of signal fading. However, due to the accumulation of noise at Alice, the scheme encounters difficulties in deriving the closed-form expressions for detection probability and optimal threshold. Building upon the CRAM scheme, Xie et al. propose a hybrid scheme based on the CR mechanism and the tags, referred to as the CR-based Hybrid (CRH) scheme \cite{CRH}. 


The motivation of this work is to address two critical limitations in existing tag-based PLA schemes. The first limitation is that the message interference caused by the decoding errors degrades the authentication performance\cite{XieSlope,PDSA}, and the second limitation is the sub-optimal threshold and sub-optimal detection probability introduced by the complexity of analyzing the decoding errors. Specifically, Bob estimates the tag by decoding the message and then removing the estimated message from the received signal. However, the imperfect decoding introduces message interference into the estimated tags in practical scenarios, and the detection performance is then significantly degraded. Moreover, due to the complexity of analyzing the decoding errors, existing schemes assume that the decoding is error-free, leading to the sub-optimal threshold and sub-optimal detection probability. 

In this paper,  we carefully design the authentication process and signal frame to propose the Tag-Based Challenge-Response (TBCR) and Series Cancellation Authentication (SCA) schemes, respectively. Specifically, the contributions of this paper are:
\begin{itemize}
	\item We propose a novel TBCR scheme that strategically benefits the strength of the CR mechanism to avoid the message interference through a carefully designed authentication process. Different from the typical PLA scheme, the tags are embedded within the forwarded challenge signal rather than directly on the message. Thus, the receiver can estimate the tag accurately by directly removing the known challenge signal. 
	\item We design the SCA scheme to eliminate message interference and address the degradation of the authentication performance caused by the accumulation of the receiver noise in the TBCR scheme. Unlike existing tag-based PLA schemes, the Series Signal Generation (SSG) and Series Signal Cancellation (SSC) modules are carefully designed for authentication signal generation and tag estimation, respectively. Thus, in the proposed SCA scheme, the tag can be estimated by superimposing series signals instead of decoding. Notably, both the TBCR and SCA schemes demonstrate distinct advantages under different practical scenarios.
	\item The closed-form expressions of the TBCR and SCA schemes are derived in terms of robustness and security. In particular, we derive the optimal threshold and detection probability regarding robustness. Our theoretical analysis reveals that the SCA scheme can achieve the ideal detection performance (say, error-free decoding) of existing tag-based schemes, while the TBCR scheme does so in the absence of noise at Alice. Furthermore, the detection probability of the eavesdropper and the key equivocation of the noisy observation are provided. 
	\item The numerical results demonstrate that the proposed theoretical results match well with the simulations, validating the effectiveness of the theoretical analysis. Meanwhile, in terms of detection probability, the SCA scheme outperforms the typical tag-based schemes, while the TBCR scheme performs better than the typical schemes at high Signal-to-Noise Ratio (SNR) regions ($>$2 dB) of Alice. In terms of security, the detection performance at Eve is significantly lower than that at Bob for both of the proposed schemes, and the TBCR scheme provides enhanced security at high SNR regions while requiring fewer keys compared with the SCA scheme. 
\end{itemize}
\vspace*{-0.1cm}

The remainder of this paper is organized as follows. Section \ref{2} introduces the system and channel models, and the limitations of existing schemes are stated. The novel TBCR and SCA schemes are proposed in Section \ref{3}, and the performance of the proposed schemes are analyzed in Section \ref{4}. In Section \ref{5}, the numerical results are carefully presented. Section \ref{6} concludes this paper.
\begin{figure}[!t]
	\setlength{\abovecaptionskip}{0pt}
	\centering
	\captionsetup{font={scriptsize}}
	\includegraphics [width=3.5in]{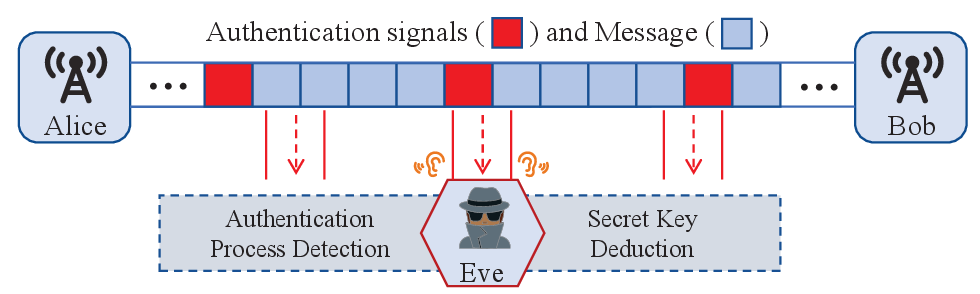}
	\caption{A typical authentication scenario.}
	\label{ThreatM}
\end{figure}

\emph{Notation}: Throughout this paper, scalars and vectors are denoted by lower-case italic letters $x$ and bold lower-case italic letters $\boldsymbol{x}$, respectively. The operator $\Re\{\cdot\}$ denotes the real part. $\boldsymbol{x}_{i}$ denotes the $i$-th bit of signal $\boldsymbol{x}$ and $\boldsymbol{x}_{i:j}$ denotes the $i$-th to $j$-th bits. $[\cdot]^*$, $[\cdot]^{\mathsf{T}}$, and $[\cdot]^{\dag}$ denote the conjugate, the transpose, and the Hermitian, respectively. The operators $\mathbb{E}(\cdot)$ and $\text{var}(\cdot)$ denotes the expectation and variance, respectively. The operators $\mathbb{H}\left(\cdot\right)$ and $||\cdot||^2$ denote the entropy and the Euclidean norm. $\text{Gen}(\cdot)$ is a hash function. $\boldsymbol{x}\sim \mathcal{CN}\left(\boldsymbol{u},\bf{\Sigma}\right)$ represents the Circularly Symmetric Complex Gaussian (CSCG) random vector $\boldsymbol{x}$ with the mean $\boldsymbol{u}$ and the covariance matrix $\bf{\Sigma}$. $\text{exp}(\cdot)$ is the exponential function. $\text{sign}(\cdot)$ is the sign function. $Q(\cdot)$ is the tail distribution function of the standard normal distribution. $\mathbb{C}^{x\times y}$ is the space of $x \times y$ complex-valued matrices. 
A permutation $P^n_k$ refers to the arrangement of $r$ elements selected from $n$ distinct elements, where the order matters. $\boldsymbol{I}_m$ is the identity matrix of order $m$.
\vspace*{-0.2cm}
\section{System Model and Typical Method}\label{2}
In this section, we first describe the system model and channel model of the tag-based PLA. Then, the typical tag-based schemes and the limitations are briefly reviewed.
\vspace*{-0.4cm}
\subsection{System Model}
We consider a typical authentication scenario as depicted in Fig. \ref{ThreatM}, where Alice transmits a tagged signal to Bob for authentication while Eve acts as an adversary. Here, Bob makes the authentication decision between the following hypotheses:
\begin{align}
	\begin{array}{l}
		{\text{H}_0}:{\rm{The\enspace received\enspace signal\enspace is\enspace  illegitimate}},\\
		{\text{H}_1}:{\rm{The\enspace received\enspace signal\enspace is\enspace legitimate}}.
	\end{array}
\end{align}
Note that the probability of incorrectly rejecting $\text{H}_0$ when $\text{H}_0$ is true is referred to as the false alarm probability, denoted by $P_{\text{FA}}$, and the probability of accepting $\text{H}_1$ when $\text{H}_1$ is true is referred to as the detection probability, denoted by $P_{\text{D}}$. The optimal detection threshold is determined by maximizing $P_{\text{D}}$ with constrained $P_{\text{FA}}$. 
Eve is a malicious adversary capable of eavesdropping. Unlike Bob, Eve is unaware of whether the legitimate parties are authenticating, but he is equipped with a powerful transceiver capable of conducting complex analyses on the received signals.
Moreover, a block fading channel is considered in the paper\cite{PLA2008,XieSlope}, where the fading coefficients remain constant during one block.
\vspace*{-0.3cm}
\subsection{Typical SUP and BTP Schemes}
In this subsection, the typical SUP and BTP schemes \cite{XiePSA,PLA2008,BTP} are briefly reviewed to clarify the limitations and to facilitate understanding the novel approaches proposed in Section \ref{3}. Note that the two typical schemes are identical in the authentication process, with the main difference being in the construction of the test statistics. In the typical schemes, tags are first embedded on the message at Alice, as illustrated in Fig. \ref{figTB}. Then, Bob decodes the message by treating the tags as interference. Finally, the tags are obtained by removing the estimated message from the received signal.

Specifically, the typical schemes comprise two phases: the preparation and authentication phases. During the preparation phase, Bob utilizes the received pilot signal to estimate the channel $\hat{h}$. In the authentication phase, Alice first embeds the tag on the message to generate the authentication signal $\boldsymbol{x}$, i.e.
\begin{align}\label{TagSup}
	{\boldsymbol{x}} = {\rho _\text{s}}{\boldsymbol{s}} + {\rho _\text{t}}{\boldsymbol{t}},
\end{align}
where $\rho_\text{s}$ and $\rho_\text{t}$ denote the power allocation of the message $\boldsymbol{s}$ and the tag $\boldsymbol{t}$, respectively, satisfying $\rho_\text{s}^2+\rho_\text{t}^2=1$ and $\rho_\text{s}\gg \rho_\text{t}$. Then, the received signal at Bob is obtained by
\begin{align}\label{eq4}
	{\boldsymbol{y}} = {h}{\boldsymbol{x}} + {\boldsymbol{w}_{\text{B}}},
\end{align}
where ${\boldsymbol{w}_{\text{B}}}$ is the receiver noise, and $h$ is the channel coefficient. Finally, Bob constructs the test statistic as	
\begin{align}
	{	\delta_{\text{SUP}}  = \Re\left\{\hat{\boldsymbol{t}}^{\dag}\boldsymbol{r}\right\}\underset{\text{H}_1}{\overset{\text{H}_0}{\lessgtr}}  \gamma_{\text{SUP}}},
\end{align}	
and
\begin{align}
	{	\delta_{\text{BTP}}  = \Re\left\{\hat{\boldsymbol{t}}^{\dag}{\hat{\boldsymbol{x}}}\right\}\underset{\text{H}_1}{\overset{\text{H}_0}{\lessgtr}}  \gamma_{\text{BTP}}},
\end{align}
where $\gamma$ and  ${\hat{\boldsymbol{x}}}$ are the threshold and the equalized signal of the received signal $\boldsymbol{y}$, respectively. $\hat{\boldsymbol{t}}$ can be expressed as
\begin{align}\label{BTPR}
	\hat{\boldsymbol{t}}=\text{Gen}\left(\hat{\boldsymbol{s}},\boldsymbol{k}_{\text{Tag}}\right),
\end{align}
and $\boldsymbol{r}$ is the tag estimated from the received signal, i.e.
\begin{align}\label{SUPr}
	\boldsymbol{r}= \frac{1}{{{\rho _\text{t}}}}({\hat{\boldsymbol{x}}} - {\rho _\text{s}}{\hat{\boldsymbol{s}}}),
\end{align}
where ${\hat{\boldsymbol{s}}}$ in (\ref{BTPR}) and (\ref{SUPr}) is the estimation of the message $\boldsymbol{s}$ by decoding.\footnote{Details about the SUP and BTP schemes can be found in \cite{PLA2008} and \cite{BTP}. Since the two schemes are similar in the authentication process, the SUP scheme is taken as an example in the following analysis.}
\begin{figure}[!t]
	\setlength{\abovecaptionskip}{0pt}
	\centering
	\captionsetup{font=scriptsize}
	\includegraphics [width=3.5in]{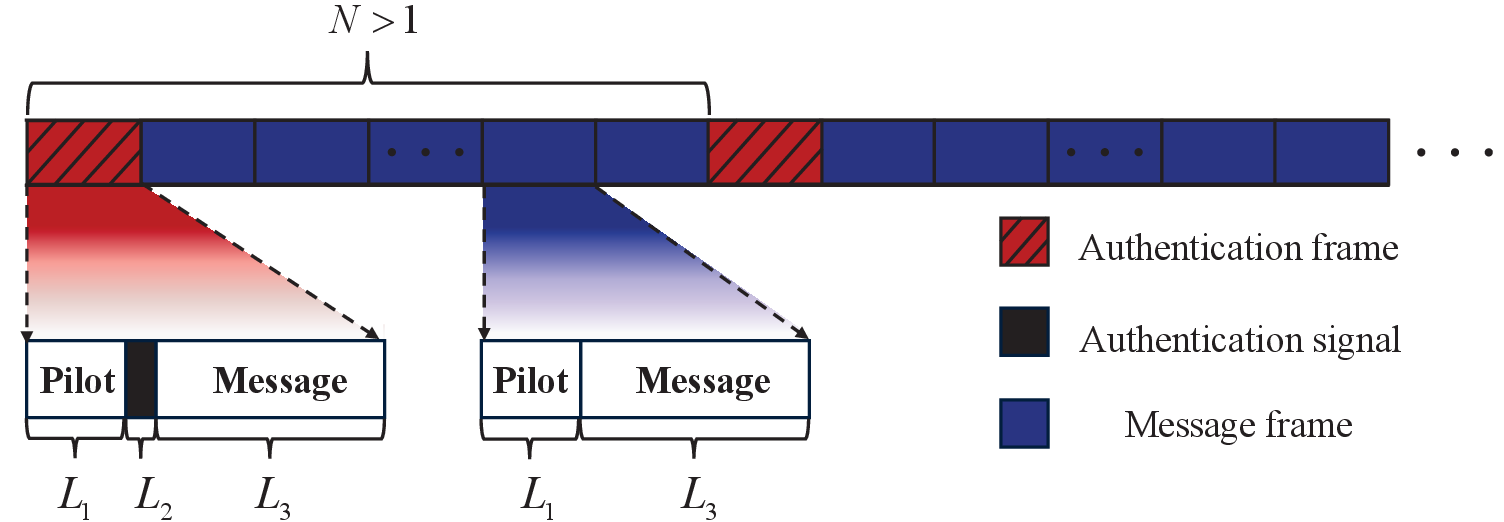}
	\caption{Frame structure of the proposed TBCR and SCA schemes.}
	\label{figFrameStru}
	\vspace*{-0.3cm}
\end{figure}
However, the typical schemes have the following two limitations: 
\begin{itemize}
	\item The first one is the message interference introduced by the decoding errors. For ease of clarity, we take the SUP scheme as an example. Specifically, the key of SUP scheme is to obtain the tag by removing the estimated message from the received signal. Unfortunately, the imperfect decoding always introduces message interference, i.e. $\hat{\boldsymbol{s}}=\boldsymbol{s}+\boldsymbol{n}_\text{D}$. Thus, the estimated tag can be given by
	\begin{align}\label{MI}
		\boldsymbol{r}=\boldsymbol{t}+\frac{\hat{h}^*}{\rho_\text{t}|\hat{h}|^2}\boldsymbol{n}-\frac{\rho_\text{s}}{\rho_\text{t}}\boldsymbol{n}_\text{D},
	\end{align}
	where the third term reduces the accuracy of the estimated tag and degrades the authentication performance.
	\item The second limitation is the sub-optimal threshold and detection probability. Due to the difficulty in deriving analytical bounds for the decoding error probability, it is assumed the decoding is error-free in the typical SUP scheme, leading to the sub-optimal threshold and detection probability\cite{DEP1,DEP2}.
\end{itemize}
 

\vspace*{-0.4cm}
\section{Proposed Physical-Layer Authentication Schemes}\label{3}
\vspace*{-0.1cm}
In this section, we first introduce the signal frame structure of the two proposed schemes. Then, both the TBCR and SCA schemes are proposed to address the message interference and sub-optimal detection probability. Compared with the typical SUP scheme, the TBCR scheme offers better detection performance and security in high SNR regions while consuming fewer keys. In contrast, the SCA scheme achieves the ideal performance in terms of the detection probability.
\vspace*{-0.4cm}

\subsection{Design of the Frame Structure}
The frame structure of the proposed schemes is shown in Fig. \ref{figFrameStru}. The authentication frames are interspersed among numerous message frames, i.e., there is one authentication frame with randomly inserted position in every $N$ message frames.  
 Within the authentication frames, the length of the pilot, authentication,
 and message signals are denoted as $L_1$, $L_2$, and $L_3$, respectively, with the condition $L_1+L_3\gg L_2$.  This frame structure offers the following advantages. First, Alice arbitrarily inserts the authentication signal frame into the message frame, reducing Eve's detection probability of the authentication process. Second, the power of the tag does not consume the power allocated to the message frame. As a result, the authentication process does not affect the demodulation of the message.
 
\begin{figure}[!t]
	\setlength{\abovecaptionskip}{0pt}
	\centering
	\captionsetup{font=scriptsize}
	\includegraphics [width=3.5in]{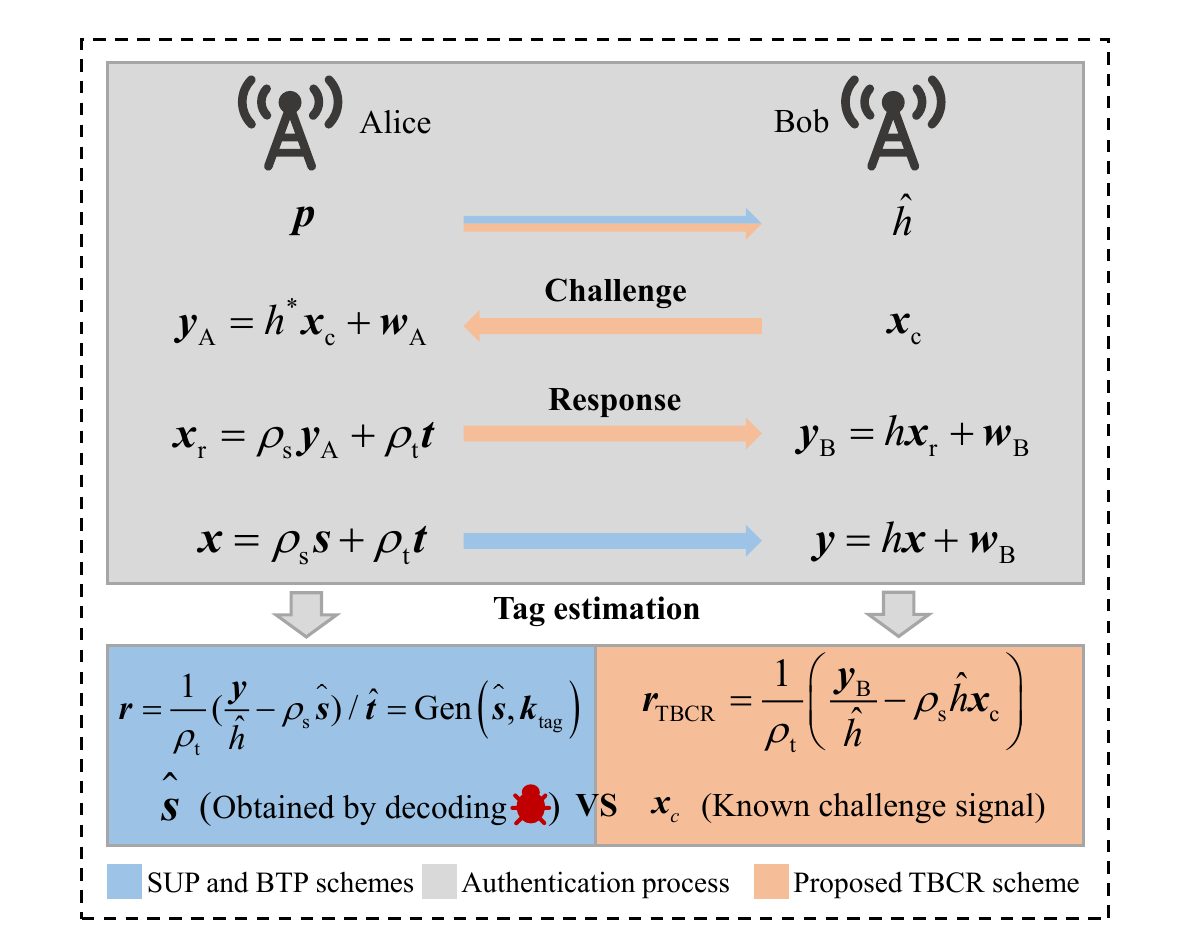}
	\caption{Authentication process and tag estimation of the SUP, BTP, and proposed TBCR schemes, where both the SUP and BTP schemes are impacted by imperfect decoding.}
	\label{figTB}
	\vspace*{-0.3cm}
\end{figure}

In the rest of this section, we will propose two authentication schemes (say, the TBCR and SCA schemes) via the designed frame structure to address the message interference.

\begin{figure*}[t]
	\centering
	\includegraphics[width=0.9\linewidth]{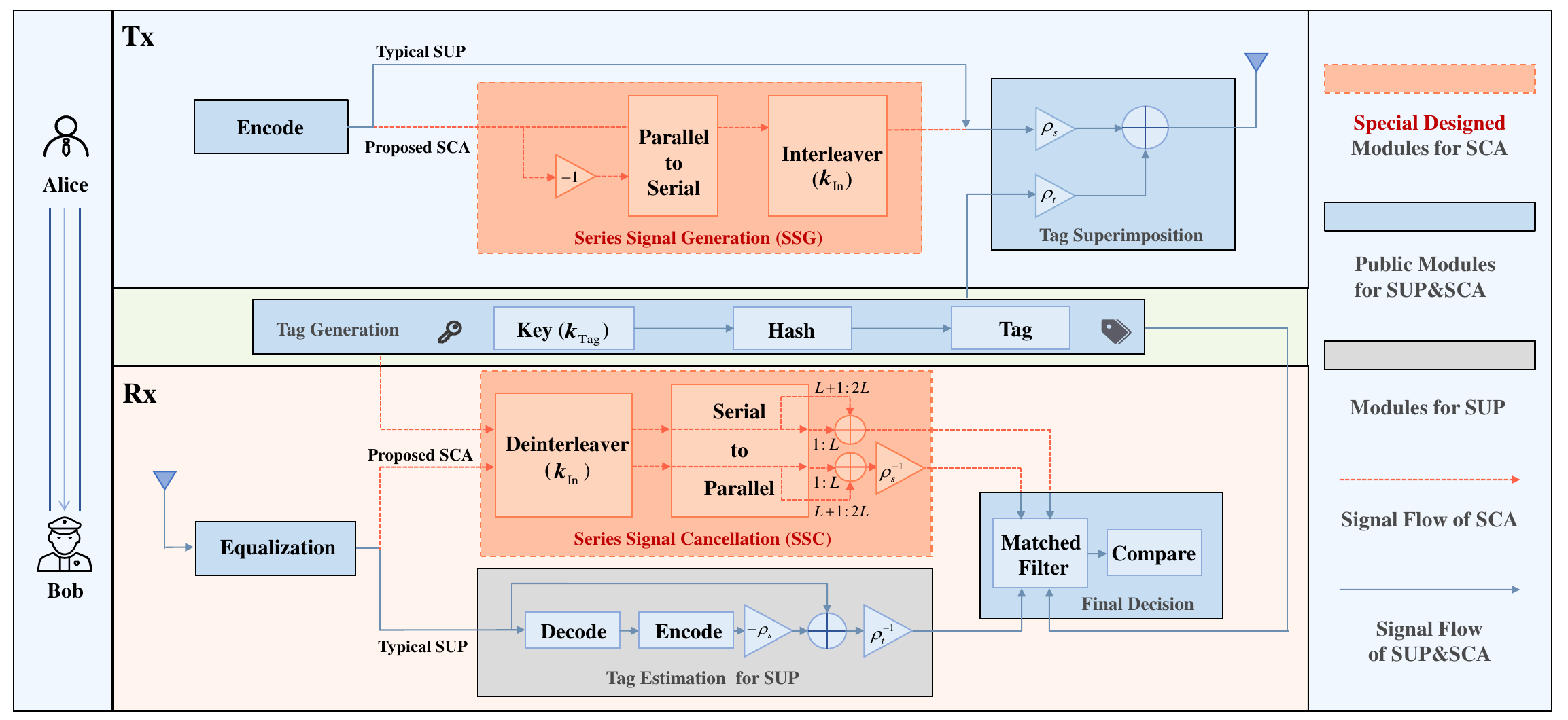} 
	\captionsetup{font=scriptsize}
	\caption{The signal flows in the authentication phase of the proposed SCA scheme versus the conventional SUP scheme, highlighting the paired modules in the SCA scheme, i.e., the novel Series Signal Generation (SSG) module at the transmitter (Alice) and the corresponding Series Signal Cancellation (SSC) module at the receiver (Bob).}
	\label{SCAandSUP}
	\vspace*{-0.4cm}
\end{figure*}
\vspace{-0.3cm}
\subsection{Proposed Tag-Based Challenge-Response Scheme}
In the CR mechanism\cite{CRAM}, Alice generates an authentication signal based on Bob's challenge signal, and Bob does not need to perform decoding for authentication. Inspired by the CR mechanism, as shown in Fig. \ref{figTB}(b), we design the TBCR scheme without decoding.

Different from the typical SUP scheme, the tag is embedded on the forwarded challenge signal in the TBCR scheme. Specifically, in the authentication phase, Alice first responses the authentication signal by superimposing the tag on the forwarded challenge signal, i.e.
	\begin{align}\label{responsesig}
	{\boldsymbol{x}_{\text{r}}} = {\rho _\text{s}}{\boldsymbol{y}_{\text{A}}} + {\rho _\text{t}}{\boldsymbol{t}},
\end{align}
where $\rho_\text{s}$ and $\rho_\text{t}$ denote the power allocation of the received challenge signal $\boldsymbol{y}_{\text{A}}\in \mathbb{C}^{L\times 1}$ and the tag $\boldsymbol{t}$. The received challenge signal $\boldsymbol{y}_{\text{A}}$ is given by
\begin{align}
	{\boldsymbol{y}_{\text{A}}} = {h^*}{\boldsymbol{x}_\text{c}} + \boldsymbol{w}_{\text{A}},
\end{align}
where $\boldsymbol{w}_{\text{A}}\sim \mathcal{CN}\left(0,\sigma_\text{A}^2\boldsymbol{I}_L\right)$ is the receiver noise and $\boldsymbol{x}_{\text{c}}$ is the challenge signal.\footnote{The proposed TBCR scheme superimposes authentication tags on the forwarded challenge signal, where the challenge signal is not restricted to specific distribution and only needs to be known by Bob. Thus, this flexibility allows the TBCR scheme to accommodate many practical signals, such as TCP acknowledgments (ACKs) and Automatic Repeat reQuest (ARQ) messages\cite{ACK}, 5G New Radio (NR) control channels signals\cite{3gpp2}, and regular signals.} The tag $\boldsymbol{t}$ is generated by
\begin{align}\label{tpk}
	\boldsymbol{t}=\text{Gen}\left(\boldsymbol{p},\boldsymbol{k}_{\text{Tag}}\right),
\end{align}
where $\boldsymbol{p}$ and $\boldsymbol{k}_{\text{Tag}}$ are the pilot signal and the secret key. Then, the received response signal at Bob is expressed by
\begin{align}\label{yb}
	{\boldsymbol{y}_{\text{B}}} = h\boldsymbol{x}_{\text{r}} + \boldsymbol{w}_{\text{B}},
\end{align}
where $\boldsymbol{w}_{\text{B}}\sim \mathcal{CN}\left(0,\sigma_\text{B}^2\boldsymbol{I}_L\right)$ is the receiver noise, and $h\sim \mathcal{CN}\left(0,\sigma_\text{h}^2\right)$ is the channel coefficient between Alice and Bob. Since Bob can estimate the tag by removing the known challenge signal from the received signal, the test statistic is finally constructed as 
\begin{align}\label{eq1}
		{\delta _{\text{TBCR}}} = \Re \left\{ \boldsymbol{t}^{\dag}{\boldsymbol{r}_{\text{TBCR}}}\right\}\underset{\text{H}_1}{\overset{\text{H}_0}{\lessgtr}}  \gamma_{\text{TBCR}},
\end{align}
where $\gamma_{\text{TBCR}}$ is the threshold determined by $P_\text{D}$ and $P_\text{FA}$, and ${\boldsymbol{r}_{\text{TBCR}}}$ is the estimated tag, i.e.
\begin{align}\label{r_TBCR}
	{\boldsymbol{r}_{\text{TBCR}}} = \frac{1}{\rho_\text{t}}\left(\frac{\boldsymbol{y}_{\text{B}}}{\hat h} - {\rho _\text{s}}{\hat{h}}{\boldsymbol{x}_\text{c}}\right).
\end{align}

\begin{remark}
	Comparing (\ref{SUPr}) and (\ref{r_TBCR}), the main difference between the TBCR scheme and the typical SUP scheme is that the decoding is used to estimate the tag in the typical SUP scheme, while the TBCR scheme can leverage the challenge signal to construct the test statistic without the additional decoding step. Thus, the TBCR scheme avoids the message interference caused by decoding errors in practical scenarios. 
\end{remark}
\vspace{-0.5cm}
\subsection{Proposed Series Cancellation Authentication Scheme}
The proposed TBCR scheme addresses the message interference in existing SUP scheme. However, the noise accumulation at Alice degrades the authentication performance. As illustrated in Fig. \ref{SCAandSUP}, a novel SCA scheme is proposed, where the paired modules, i.e., SSG and SSC, are carefully designed to generate the series signal and estimate the tag, respectively. Since the tag is embedded on the well-designed series signal, it can be accurately estimated through a folding strategy. 

\textbf{Series Signal Generation Module}: The SSG module incorporates both the parallel-to-serial and interleaver modules. First, we generate series signals by splicing two sets of signals that are identical in content but opposite in signs using the multiplier and parallel-to-serial modules. The core motivation of the series signal is to enable Bob to cancel out the message without decoding when estimating the tags. Specifically, the encoded signal $\boldsymbol{m}\in\mathbb{C}^{L\times1}$
is multiplied by $-1$ to obtain the signal $-\boldsymbol{m}$. After passing through the parallel-to-serial module, the spliced signal $[\boldsymbol{m}^{\mathsf{T}},-\boldsymbol{m}^{\mathsf{T}}]^{\mathsf{T}}$ with length $2L$ is generated. To prevent easy detection by Eve, the spliced signal is then passed through the interleaver, resulting in the chaotic signal $\boldsymbol{s}$, i.e.
\begin{align}
\boldsymbol{s} = \ell\left([\boldsymbol{m}^{\mathsf{T}},-\boldsymbol{m}^{\mathsf{T}}]^{\mathsf{T}}\right),
\end{align}
where $\ell\left(\cdot\right)$ is the interleaving, determined by the secret key $\boldsymbol{k}_{\text{in}}$.\footnote{For example, if $L=3$, $\ell=[5,3,4,1,2,6]^\mathsf{T}$, and $\boldsymbol{m}=[a,b,c]^{\mathsf{T}}$, then we have $\boldsymbol{s}=[-b,c,-a,a,b,-c]^\mathsf{T}$.} Finally, the tag generated by (\ref{tpk}) is superimposed on the chaotic signal $\boldsymbol{s}$ as in (\ref{TagSup}) to generate the authentication signal. 

\textbf{Series Signal Cancellation Module}: The deinterleaver and serial-to-parallel modules are included in the SSC module. At the receiver, Bob can estimate the tag by directly folding the series signal instead of decoding. Specifically, the equalized signal $\hat{\boldsymbol{s}}$ at Bob is expressed as 
\begin{align}\label{eqss}
	\hat{\boldsymbol{s}} = {\rho _\text{s}}{\boldsymbol{s}} + {\rho _\text{t}}{\boldsymbol{t}} + \frac{\boldsymbol{w}_\text{B}}{\hat h},
\end{align}
where $\boldsymbol{w}_{\text{B}}\sim \mathcal{CN}\left(0,\sigma_\text{B}^2\boldsymbol{I}_{2L}\right)$ is the receiver noise. Corresponding to SSG module, the equalized signal is first passed through the deinterleaver to obtain 
	\begin{align}\label{hatm+}
	{\hat{\boldsymbol{m}} } =\rho_\text{s}\ell^{-1}(\boldsymbol{s}) + {\rho _\text{t}}\boldsymbol{t}^{\text{DE}} + \frac{\boldsymbol{w}_{\text{B}}^{\text{DE}}}{{\hat h}},
\end{align}
where $\ell^{-1}(\cdot)$ is the reverse mapping of $\ell(\cdot)$; $\boldsymbol{w}_{\text{B}}^{\text{DE}}=\ell^{-1}(\boldsymbol{w}_\text{B})$; $\ell^{-1}(\boldsymbol{s}) =[\boldsymbol{m}^{\mathsf{T}},-{\boldsymbol{m}}^{\mathsf{T}}]^{\mathsf{T}}$; $\boldsymbol{t}^{\text{DE}}=\ell^{-1}(\boldsymbol{t})$. We can observe that the first $L$ bits and the last $L$ bits of the first item in (\ref{hatm+}) have the same content but the opposite signs. Therefore, a serial-to-parallel module can be used to obtain $\hat{\boldsymbol{m}}_{1:L}$ and $\hat{\boldsymbol{m}}_{L+1:2L}$, respectively. Subsequently, the estimation of the tag is obtained by folding the signals $\hat{\boldsymbol{m}}_{1:L}$ and $\hat{\boldsymbol{m}}_{L+1:2L}$, i.e.
\begin{align}\label{SCAR}
	\boldsymbol{r}_{\text{SCA}} =\frac{1}{{\rho _\text{t}}} ({\hat{\boldsymbol{m}}_{1:L} } + \hat{\boldsymbol{m}}_{L+1:2L} ) = \boldsymbol{T} + \frac{{\boldsymbol{N}_\text{B}}}{({\rho _\text{t}}{\hat h})},
\end{align}
where 
\begin{align}
	\boldsymbol{T}=\boldsymbol{t}^{\text{DE}}_{1:L} + \boldsymbol{t}^{\text{DE}}_{L+1:2L},
\end{align}
and
\begin{align}
	\boldsymbol{N}_{\text{B}}=\boldsymbol{w}^{\text{DE}}_{\text{B},1:L} + \boldsymbol{w}^{\text{DE}}_{\text{B},L+1:2L}.
\end{align}
Note that the second term in (\ref{eqss}) also undergoes the SSC module. To detect the presence of the tag, an initial tag $\boldsymbol{t}$ is generated using the secret key $\boldsymbol{k}_{\text{Tag}}$, and $\boldsymbol{T}$ is obtained after subjecting the initial tag $\boldsymbol{t}$ to the SSC module. At last, a test statistic is constructed as
\begin{align}\label{delta_SCA}
	{\delta _{\text{SCA}}} = \Re \left\{ \frac{1}{2}\boldsymbol{r}_{\text{SCA}}^{\dag}{\boldsymbol{T}}\right\}\underset{\text{H}_1}{\overset{\text{H}_0}{\lessgtr}}  \gamma_{\text{SCA}},
\end{align}
where $1/2$ is a normalized coefficient. It can be observed that the SCA scheme ingeniously designs the signals frame, enabling the receiver to obtain an estimate of the tag through a folding strategy, rather than the decoding process. This enhances the estimation accuracy of the tag and simplifies the complexity at the receiver.

\begin{remark}
Here, we compare the SCA scheme with the SUP and TBCR schemes. In the typical SUP scheme, the message interference is introduced by imperfect decoding. In the proposed SCA scheme, the tag is estimated by folding two signals rather than decoding. Thus, different from the SUP scheme, the SCA scheme avoids the decoding errors by carefully designing the series signal. Comparing (\ref{yb}) and (\ref{eqss}), the main difference between the SCA and TBCR schemes is that the extra noise, which degrades the detection performance, is introduced in the TBCR scheme. For the SCA scheme, no additional noise is introduced. However, the key consumption of the SCA scheme is twice that of the TBCR scheme.
\end{remark}
\vspace*{-0.3cm}
\section{Performance Analysis}\label{4}
In this section, we analyze the performance of the proposed schemes. Specifically, the closed-form expressions for Bob's optimal detection threshold and detection probability under different SNRs are derived, demonstrating the robustness of the proposed schemes to both noise and channel fading. In terms of security, we analyze Eve's detection of the authentication behavior and the difficulty of Eve deciphering tags. Moreover, we further discuss the proposed schemes in terms of Ratio of Bandwidth Efficiency (RBE), time complexity, and applicability.
\vspace*{-0.2cm}
\subsection{Robustness Analysis}
The robustness of the PLA scheme, which indicates the ability of Bob to distinguish legitimate signals from noisy observations, is measured by the false alarm probability and the detection probability\cite{CRH}. 
\subsubsection{Robustness Analysis for the TBCR Scheme}
The closed-form expressions of the robustness for the TBCR scheme are provided in the following proposition.
\begin{proposition}\label{pro1}
	In the TBCR scheme, the closed-form expressions of the false alarm probability and the detection probability at Bob are given by (\ref{eqTBCR_PFA}) and (\ref{eqTBCR_PD}), respectively, shown at the top of the next page. The optimal threshold can be obtained through traversal of (\ref{eqTBCR_PFA}). Furthermore, in the high SNR regions of Alice, i.e., $\sigma_{\text{A}}^2 \rightarrow 0$, the false alarm probability, the optimal threshold, and the detection probability for the TBCR scheme can be expressed as
\begin{figure*}[!t]
		\begin{align}\label{eqTBCR_PFA}
			{P_{\text{FA},\text{TBCR}}} \approx \frac{1}{2}\left(1 - \sqrt {\frac{a}{{\sigma _\text{A}^2}}} \frac{\pi }{N}\sum\limits_{i = 1}^N {\sqrt {1 - \cos^2\left(\frac{{(2i - 1)\pi }}{{2N}}\right)} } {f_{\text{FA}}}\left(\frac{1}{2}\sqrt {\frac{a}{{\sigma _\text{A}^2}}} \cos\left(\frac{{(2i - 1)\pi}}{{2N}}\right) + \frac{1}{2}\sqrt {\frac{a}{{\sigma _\text{A}^2}}} \right)\right).
		\end{align}
		\begin{align}\label{eqTBCR_PD}
			{P_{\text{D},\text{TBCR}}} \!\approx\! \frac{1}{2}\left(\!1\! - \!\text{sign}\!\left({\gamma} - L\right)\!\sqrt {\frac{b}{{\rho _\text{s}^2\sigma _\text{A}^2}}} \frac{\pi }{N}\sum\limits_{i = 1}^N {\sqrt {1 - \cos^2\left(\frac{{\left(2i - 1\right)\pi }}{{2N}}\right)} } {f_\text{D}}\!\left(\frac{1}{2}\sqrt {\frac{b}{{\rho _\text{s}^2\sigma _\text{A}^2}}} \cos\!\left(\!\frac{{\left(2i - 1\right)\pi }}{{2N}}\!\right) \!+\! \frac{1}{2}\sqrt {\frac{b}{{\rho _\text{s}^2\sigma _\text{A}^2}}}\right)\!\right).
		\end{align}
		\hrulefill \vspace*{-10pt}
\end{figure*} \leavevmode \\
\begin{align}\label{TBCRNoAlice1}
		{P_{\text{FA},\text{TBCR},\text{Ideal}}} = \frac{1}{2}\left(1 - \sqrt {\frac{{{\gamma ^2}\rho_\text{t}^2\sigma_\text{h}^2}}{{L\sigma_\text{B}^2 + {\gamma ^2}\rho_\text{t}^2\sigma_\text{h}^2}}} \right),
\end{align}
\begin{align}\label{TBCRNoAlice2}
		{\gamma _{\text{TBCR},\text{Ideal}}} = (1 - 2\epsilon)\sqrt {\frac{{L\sigma_\text{B}^2}}{{4\epsilon(1 - \epsilon)\rho_\text{t}^2\sigma_\text{h}^2}}}, 
\end{align}
	and
\begin{align}\label{TBCRNoAlice3}
		{P_{\text{D},\!\text{TBCR},\text{Ideal}}} \!=\! \frac{1}{2}\!\left(\! \!{1\!\! -\!\! \text{sign}({\gamma}\! -\! L)}\! \sqrt {\!\frac{{\rho_\text{t}^2\sigma_\text{h}^2{{({\gamma}\! -\! L)}^2}}}{{L\sigma_\text{B}^2 \!+\!\rho_\text{t}^2\sigma_\text{h}^2{{({\gamma}\! -\! L)}^2}}}\!} \right)\!.
\end{align}
	In (\ref{eqTBCR_PFA}), $\gamma$ is a given threshold, $a = 2{\gamma ^2}\rho _\text{t}^2/L$, and ${f_{\text{FA}}}(x) = \frac{1}{{\sqrt {2\pi } }}{\text{exp}( - \frac{{\sigma _\text{B}^2{x^2}}}{{\sigma _\text{h}^2(a - \sigma _\text{A}^2{x^2})}} - \frac{{{x^2}}}{2})}$. In (\ref{eqTBCR_PD}), ${f_\text{D}}(x) = \frac{1}{{\sqrt {2\pi } }}{\text{exp}( - \frac{{\sigma _\text{B}^2{x^2}}}{{\sigma _\text{h}^2(b - \rho _\text{s}^2\sigma _\text{A}^2{x^2})}} - \frac{{{x^2}}}{2})}$, $b = 2\rho _\text{t}^2{({\gamma} - L)^2}/L$, and $N$ is a complexity-accuracy tradeoff parameter.
\end{proposition}

\begin{proof_black}
	See Appendix A.
\end{proof_black}
\vspace*{0.2cm}
\subsubsection{Robustness Analysis for the SCA Scheme}
The closed-form expressions of the SCA scheme in terms of robustness are presented through the following proposition.
\begin{proposition}
	In the SCA scheme, the closed-form expressions of the false alarm probability, the optimal threshold, and the detection probability at Bob can be respectively given by 
	\begin{align}
		{P_{\text{FA},\text{SCA}}} = \frac{1}{2}\left(1 - \sqrt {\frac{{{\gamma ^2}\rho _\text{t}^2\sigma _\text{h}^2}}{{L\sigma _\text{B}^2 + {\gamma ^2}\rho _\text{t}^2\sigma _\text{h}^2}}} \right),
	\end{align}
	\begin{align}\label{gamma_SCA}
		{\gamma _{\text{SCA}}} = (1 - 2\epsilon)\sqrt {\frac{{L\sigma _\text{B}^2}}{{4\epsilon(1 - \epsilon)\rho _\text{t}^2\sigma _\text{h}^2}}}, 
	\end{align}
	and
	\begin{align}\label{PD_SCA}
		P_{\text{D},\text{SCA}}=&\frac{1}{2}\Bigg(1-\text{sign}(\gamma-L)\notag\\
		&\left.\times\sqrt{\frac{\rho_{t}^{2}\sigma_{\text{h}}^{2}(\gamma-L)^{2}}{L\sigma_{\text{B}}^{2}+\rho_{\text{t}}^{2}\sigma_{\text{h}}^{2}(\gamma-L)^{2}}}\right),
	\end{align}
	where $\gamma$ is a given threshold and $\epsilon$ is the upper limit of the false alarm probability.
\end{proposition}
\begin{proof_black}
	See Appendix B. 
\end{proof_black}

\begin{remark}
The relationship of the detection performance between the TBCR and SCA schemes can be expressed as
\begin{align}
	P_{\text{FA},\text{SCA}}=P_{\text{FA},\!\text{TBCR},\text{Ideal}}=P_{\text{FA},\text{SUP},\text{Ideal}},
\end{align}
\begin{align}
		\gamma _{\text{SCA}}=\gamma _{\text{TBCR},\text{Ideal}}=\gamma _{\text{SUP},\text{Ideal}},
\end{align}
and
\begin{align}
	P_{\text{D},\text{SCA}}=P_{\text{D},\!\text{TBCR},\text{Ideal}}=P_{\text{D},\text{SUP},\text{Ideal}}\geq P_{\text{D},\!\text{TBCR}},
\end{align}
where $P_{\text{FA},\text{SUP},\text{Ideal}}$, $\gamma _{\text{SUP},\text{Ideal}}$, and $P_{\text{D},\text{SUP},\text{Ideal}}$ are closed-form expressions of the SUP scheme in terms of the false alarm probability, optimal threshold, and detection probability under the assumption of error-free decoding, given by in (40), (41), and (44) in \cite{XiePSA}. It is observed that the SCA scheme achieves the ideal performance of the SUP scheme, while the TBCR scheme approaches the ideal performance as the SNR at Alice increases. In practical scenarios, the performance comparison between the TBCR and SUP schemes depends on the SNR at Alice, which will be provided in Section \ref{5}.
\end{remark}

\subsubsection{Non-Ideal Conditions}
In this subsection, we analyze the proposed schemes under non-ideal conditions. For example, the perfect channel estimation is inherently challenging to obtain due to factors such as device measurement errors and receiver noise. Consequently, the Channel State Information (CSI) errors are introduced, i.e.
\begin{align}\label{IMCSI}
	h = \hat{h}+\triangle h,
\end{align}
where the channel $h \sim \mathcal{CN}\left(0,\sigma_\text{h}^2\right)$ and the CSI error $\triangle h\sim \mathcal{CN}\left(0,\sigma_\text{e}^2\right)$ are assumed to follow the CSCG distribution. For the statistical CSI error model, the variance of $\triangle h$ is defined as $\sigma_\text{e}^2=\eta_\text{e}^2\sigma_\text{h}^2$, where the normalized CSI error $\eta_\text{e}^2\in [0,1)$ measures the relative amount of CSI uncertainties\cite{ImCSI1}.

For the TBCR scheme, the estimation of the tag with channel estimation errors under $\text{H}_1$ can be expressed as 
\begin{align}
{{\boldsymbol{r}}_{{\text{TBCRe|}}{{\text{H}}_1}}} =& \frac{{\Delta {h^2}{\rho _s}{{\boldsymbol{x}}_c} + {{\boldsymbol{w}}_{\text{B}}}}}{{{\rho _t}\hat h}}\notag\\
&+\!\left( {1 \!+\! e} \right){\boldsymbol{t}}\! +\! \frac{1}{{{\rho _t}}}\left( {\left( {1\! +\! e} \right){\rho _s}{{\boldsymbol{w}}_{\text{A}}}\! +\! 2\Delta h{\rho _s}{{\boldsymbol{x}}_c}} \right),	
\end{align}
where $e = {\Delta h}/{\hat h}$ is the relative error. Thus, the test statistic in (\ref{eq1}) follows
\begin{align}\label{TBCRerrH1}
{\delta _{{\text{TBCR|}}{{\text{H}}_1}}}\sim\mathcal{CN}\left( {L,\frac{{L\left( {\rho _s^2\sigma _{\text{A}}^2|\hat h{|^2} + \sigma _{{\text{TBCRe}}}^2} \right)}}{{2\rho _t^2|\hat h{|^2}}}} \right),
\end{align}
where $\sigma _{{\text{TBCRe}}}^2 = \sigma _{\text{B}}^2 + \left( {L\rho _t^2 + \rho _{\text{s}}^2\sigma _{\text{A}}^2} \right)\eta _{\text{e}}^2\sigma _{\text{h}}^2$ is the cumulative error in the TBCR scheme. Compared to (\ref{delta_TBCR_H1}), $\eta _{\text{e}}^2$ leads to an increase in the noise, which in turn raises the variance of the test statistic. Thus, the uncertainty in detection increases. Moreover, a comparison of (\ref{TBCRerrH1}) and (\ref{delta_TBCR_H1}) reveals that the detection probability with  channel estimation errors can be readily derived by substituting $\sigma_{\text{B}}^2$ in (\ref{eqTBCR_PD}) with $\sigma _{{\text{TBCRe}}}^2$. 

Similarly, for the SCA scheme, the estimation of the tag with channel estimation errors under $\text{H}_1$ can be expressed as
\begin{align}
{{\boldsymbol{r}}_{{\text{SCAe|}}{{\text{H}}_1}}} = \left( {1 + e} \right)T + \frac{{{{\boldsymbol{N}}_{\text{B}}}}}{{\left( {{\rho _t}\hat h} \right)}}.
\end{align}
According to (\ref{delta_SCA}), the test statistic with the channel estimation errors follows
\begin{align}
{\delta _{{\text{SCA|}}{{\text{H}}_1}}}\sim\mathcal{CN}\left( {L,\frac{{L\sigma _{{\text{SCAe}}}^2}}{{2\rho _t^2|\hat h{|^2}}}} \right),
\end{align}
where $\sigma _{{\text{SCAe}}}^2 = \sigma _{\text{B}}^2 + \rho _{\text{t}}^2L\eta _{\text{e}}^2\sigma _{\text{h}}^2$ is the cumulative error in the SCA scheme. Thus, for the SCA scheme, the detection probability with the channel estimation error can be achieved by substituting $\sigma_{\text{B}}^2$ in (\ref{PD_SCA}) with $\sigma _{{\text{SCAe}}}^2$. Since $\sigma _{{\text{TBCRe}}}^2 >\sigma _{{\text{SCAe}}}^2$ holds,  the impact of channel estimation errors on the SCA scheme is smaller than that on the TBCR scheme.

\vspace{-0.4cm}
\subsection{Security Analysis}
\vspace{-0.1cm}
In this subsection, the proposed schemes are analyzed from the perspective of the adversary. Our security analysis follows the traditional PLA threat model, where Eve knows all system parameters except the shared secret key and can intercept the challenge and response signals in the TBCR scheme. 

\subsubsection{Eve's Detection}
In the authentication process, Alice selectively inserts the tag into the message to confuse Eve. As an adversary, Eve may first observe whether the authentication process occurs from the noisy observation. Unlike Bob, Eve does not know the key and can not directly detect the tags in the signal. However, since the tagged signals exhibit different statistical distribution compared to message, Eve can infer the authentication process based on the signal distribution. In this subsection, we analyze Eve's detection of the authentication process in the TBCR and SCA schemes, respectively.

\textbf{Eve's Detection in the TBCR Scheme:} In the TBCR scheme, a capable Eve can eavesdrop on both the challenge signal $\boldsymbol{x}_\text{c}$ and the response signal $\boldsymbol{x}_\text{r}$, and then remove $\boldsymbol{x}_\text{c}$ from the received response to estimate the tag. To demonstrate the security of the proposed schemes, we consider two worst cases for the legitimate parties, which are (i) Eve is close to Alice, and (ii) Eve is close to Bob. The upper bound for the detection probability of Eve in high SNR regions is given in the following proposition.

\begin{proposition}
	In the TBCR scheme, the closed-form expressions of the false alarm probability and the optimal threshold of Eve when he is close to Alice can be respectively given by
	\vspace{-0.1cm}
	\begin{align}
		{P_{\text{FA},\text{TBCR,EA}}} = \frac{1}{2}\left(1 - \sqrt {\frac{{{\gamma ^2}\sigma_\text{h}^2}}{{(\sigma_\text{A}^2 + \sigma_\text{E}^2)L + {\gamma ^2}\sigma_\text{h}^2}}} \right),
	\end{align}
	and
	\begin{align}
		{\gamma _{\text{TBCR,EA}}} = \sqrt {\frac{{{{\left(1 - 2\epsilon\right)}^2}\left(\sigma_\text{A}^2 + \sigma_\text{E}^2\right)L}}{{4\epsilon\left(1 - \epsilon\right)\sigma_\text{h}^2}}},
	\end{align}
	and the upper bound of the detection probability in high SNR regions is given by
		\begin{align}\label{TBCR_EA}
		P_{\text{D},\text{TBCR,EA}}=&\frac{1}{2}\Bigg(1-\text{sign}\left(\gamma-\left(1-\rho_{\text{s}}\right)L\right)\notag\\
		&\left.\times\sqrt{\frac{\left(\gamma\!\!-\!\!\left(1\!\!-\!\!\rho_{\text{s}}\right)L\right)^{2}\sigma_{\text{h}}^{2}}{\!\sigma_{\text{E}}^{2}L\!+\!\left(\gamma\!-\!\!\left(1\!\!-\!\!\rho_{\text{s}}\right)\!L\right)^{2}\!\sigma_{\text{h}}^{2}}}\right),
	\end{align}
	 When Eve is close to Bob, corresponding closed-form expressions and the upper bound can be expressed as
	\begin{align}
		{P_{\text{FA},\text{TBCR,EB}}} = \frac{1}{2}\left(1-2g\left(\sqrt{\frac{L\sigma_\text{E}^2}{2\gamma^2\sigma_{\text{h}}^4}}\right)\right),
	\end{align}
	\begin{align}
		{\gamma _{\text{TBCR,EB}}} = \sqrt{\frac{L\sigma_\text{E}^2}{2\left(g^{-1}\left(0.5-\epsilon\right)\right)^2\sigma_{\text{h}}^4}},
	\end{align}
	and
	\begin{align}\label{TBCR_EB}
		\!P_{\text{D},\text{TBCR,EB}}\!=\!\frac{1}{2}\!\left(1\!-\!2{\text{sign}}\left(\gamma\!-\!\left(1\!-\!\rho_{\text{s}}\right)L\right)g\left(\sqrt{2K_\gamma}\right)\right),
	\end{align}
	where $g(x)=\exp({x^2/2})Q(x)$ is a monotonic function with $x\in(0,+\infty)$, and $g^{-1}(\cdot)$ is the inverse function of $g(\cdot)$. $\epsilon$ is the maximum of $P_{\text{FA}}$, and $K_{\gamma}=\frac{L\sigma_{\text{E}}^{2}}{4(\gamma-(1-\rho_{\text{s}})L)^{2}\sigma_{\text{h}}^{4}}$ holds.
	
\end{proposition}

\begin{proof_black}
		See Appendix C. 
\end{proof_black}
\begin{remark}
For the TBCR scheme, the upper bound of Eve's performance in high SNR regions (corresponding to an SNR where the detection probability exceeds 50\%) are derived in (\ref{TBCR_EA}) and (\ref{TBCR_EB}) . Although Eve can eavesdrop on both the challenge and response signals, this upper bound remains significantly lower than Bob's detection performance through numerical results, indicating the security of the TBCR scheme.
\end{remark}

\textbf{Eve's Detection in the SCA Scheme:} In the SCA scheme, Eve first determines whether authentication is present in the current communication. However, it is challenging for Eve to make an accurate decision without the knowledge of the secret keys. Thus, he may first estimate the signal $\boldsymbol{s}$. Under the cover of the noise and tags, Eve can estimate the signal $\boldsymbol{s}$ through two methods. (i) A hard decision is made directly on the received signal, that is, Eve directly obtains the signal $\boldsymbol{s}$ through demodulation. However, there is still a non-negligible bit error rate (BER) even when the SNR is 10 dB for the BPSK signal and the demodulation SNR is further reduced when a tag is superimposed. (ii) Like the SUP scheme, Eve first demodulates and decodes the signal, followed by re-encoding and re-modulation to obtain signal $\boldsymbol{s}$. For Eve to decode, he must first guess the interleaver without the secret key $\boldsymbol{k}_\text{In}$. When the length of the encoded signal and the interleaver are $L$ and $2L$ respectively, the probability of Eve accurately estimating the interleaver is $1/P^{2L}_{2L}$. After each estimation of the interleaver, Eve can verify this by decoding the first and last $L$ symbols separately of the deinterleaved signal and comparing the results of the two parts. Here, a powerful computational capabilities is assumed for Eve. It is worthy noting that in practical scenarios, even with Eve's accurate estimations of the interleaver, the security of the SCA scheme is still the same as that of the typical tag-based schemes. 

The received signal at Eve is equalized based on ${h}_{\text{EA}}$ as
\begin{align}
	{\hat{\boldsymbol{s}}_\text{E}} = {\rho _\text{s}}{\boldsymbol{s}} + {\rho _\text{t}}{\boldsymbol{t}} +\frac{{\boldsymbol{w}_\text{E}}}{{h_\text{EA}}},
\end{align}
where $h_{\text{EA}}\sim \mathcal{CN}\left(0,\sigma_{\text{EA}}^2\right)$. Then,  deinterleaving is performed on signal $\hat{\boldsymbol{s}}_\text{E}$ based on the secret key $\boldsymbol{k}_\text{In}$ to obtain signal
\begin{align}\label{mmm}
	\hat{\boldsymbol{m}}_{\text{E}}  = {\rho _\text{s}}\ell^{-1}(\boldsymbol{s}) + {\rho _\text{t}}{\ell ^{ - 1}}({\boldsymbol{t}}) +\frac{\ell ^{ - 1}({\boldsymbol{w}_\text{E}})}{{h_\text{EA}}}.
\end{align}
However, Eve is unable to perform matched filtering on the tag without the secret key $\boldsymbol{k}_{\text{Tag}}$, and he can only use the signal $\boldsymbol{m}$ to construct the test statistic as
\begin{align}
	{\delta _{\text{E,SCA}}}  = \Re\left\{ {\left(\boldsymbol{m} - \hat{\boldsymbol{m}}_{\text{E},1:L}\right)^{\dag}}\boldsymbol{m}\right\}.
\end{align}
The test statistic under different hypotheses can be given by
\begin{align}\label{SCA_E_H0}
	{\delta _{\text{E,SCA}|{\text{H}_0}}}\sim \mathcal{CN}\left(0,\frac{L\sigma _\text{E}^2}{|2{h_\text{EA}}{|^2}}\right),
\end{align}
\vspace{-0.1cm}
and
\vspace{-0.2cm}
\begin{align}\label{SCA_E_H1}
	{\delta _{\text{E,SCA}|{\text{H}_1}}} \sim \mathcal{CN}\left((1 - {\rho _\text{s}})L,\frac{L\sigma _\text{E}^2}{|2{h_\text{EA}}{|^2}}\right).
\end{align}

The optimal threshold for Eve's detection in the SCA scheme is proposed, as well as the false alarm probability and detection probability through the following proposition.
\begin{proposition}
	In the SCA scheme,  the false alarm probability, the optimal threshold, and the detection probability at Eve can be respectively given by
	\begin{align}
		{P_{\text{FA},\text{E,SCA}}} = \frac{1}{2}\left(1 - \sqrt {\frac{{{\gamma ^2}\sigma _{\text{EA}}^2}}{{L\sigma _\text{E}^2 + {\gamma ^2}\sigma _{\text{EA}}^2}}} \right),
	\end{align}
	\begin{align}
		{\gamma _\text{E,SCA}} = \left(1 - 2{\epsilon}\right)\sqrt {\frac{{L\sigma _\text{E}^2}}{{4{\epsilon}(1 - {\epsilon})\sigma _{\text{EA}}^2}}} ,
	\end{align}
and
	\begin{align}\label{SCA_E}
		P_{\text{D},\text{E,SCA}}=&\frac{1}{2}\Bigg(1-\text{sign}\left(\gamma-(1-\rho_{\text{s}})L\right)\notag\\
		&\left.\times\sqrt{\frac{\sigma_{\text{EA}}^{2}\left(\gamma-\left(1-\rho_{\text{s}}\right)L\right)^{2}}{L\sigma_{\text{E}}^{2}+\sigma_{\text{EA}}^{2}\left(\gamma-\left(1-\rho_{\text{s}}\right)L\right)^{2}}}\right).
	\end{align}
\end{proposition}
\vspace*{0.2cm}

\begin{proof_black}
	The proof is analogous to that of Proposition 1.
\end{proof_black}

\begin{remark}
	 Eve's detection performance in (\ref{SCA_E}) remains inferior to Bob's in the numerical results despite a powerful computational capabilities of Eve, further attesting to the superiority of the SCA scheme. This reveals the key characteristic of PLA, that is, even powerful computing capabilities cannot guarantee the cracking of authentication.
\end{remark}

\subsubsection{Key Equivocation}
The key equivocation measures the difficulty for Eve to obtain the tag in the noisy observation from the information theory perspective \cite{PLA2008}. The higher the key equivocation, the greater the security. In this subsection, the Tag-to-Noise Ratio (TNR) of the TBCR and SCA schemes is derived first, and the key equivocation is given further.

In the typical SUP scheme, the tag is protected solely by the receiver noise and message as in (\ref{eq4}). In contrast, the tag is protected by the noise at Alice and Eve, and challenge signal $\boldsymbol{x}_c$ in the TBCR scheme as in (\ref{yb}). Thus, the noise at Alice facilitates the concealment of the low-power tag. The TNR in the TBCR scheme can be expressed as
\begin{align}
	{\vartheta _{\text{TBCR}}} = \frac{{\rho_\text{t}^2|{h_{\text{EA}}}{|^2}}}{{\rho_\text{s}^2|{h_{\text{EA}}}{|^2}\sigma_\text{A}^2 + \sigma_\text{E}^2}}.
\end{align}
In the SCA scheme, Eve estimates the tag with his powerful computational capabilities to obtain
\begin{align}
	{\hat{\boldsymbol{t}}_\text{E}} = \boldsymbol{T} + \frac{\boldsymbol{w}_{\text{E,de}}^{1:L} + \boldsymbol{w}_{\text{E,de}}^{L+1:2L}}{{\rho _\text{t}}{h_{\text{EA}}}}.
\end{align}
Thus, the TNR in the SCA scheme can be expressed as
\begin{align}
	{\vartheta _{\text{SCA}}} = \frac{{\rho _\text{t}^2|{h_{\text{EA}}}{|^2}}}{{\sigma _\text{E}^2}}.
\end{align}
The key equivocation of the TBCR and SCA schemes is given by the following proposition.

\begin{proposition}
The closed-form expression of the key equivocation of the TBCR and SCA schemes can be expressed as
\begin{align}\label{TBCRKE}
	\mathbb{H}(\boldsymbol{t}|{\boldsymbol{y}_\text{E}}) = {P_e}\text{log}_2\frac{1}{{{P_e}}} + \left(1 - {P_e}\right)\text{log}_2\frac{1}{{1 - {P_e}}},
\end{align}
where ${P_e} = Q(\sqrt {{\mathbb{E}}({\vartheta})} )$, $\vartheta=\vartheta_{\text{TBCR}}$ for the TBCR scheme, and $\vartheta=\vartheta_{\text{SCA}}$ for the SCA scheme.
\end{proposition}

\subsubsection{Other Attacks}

Similar to typical SUP scheme, the proposed schemes may be susceptible to replay attacks. Bob can distinguish the attack by analyzing the noise distribution of the received signal\cite{AgainstReply}. 

For the TBCR scheme, Eve initiates the replay attacks by forwarding Alice's response signal. The received signal at Bob can be expressed as
\begin{align}
{{\boldsymbol{y}}_{{\text{TBCR|Eve}}}}=& {h_{{\text{EB}}}}\left( {{h_{{\text{AE}}}}\left( {{\rho _{\text{s}}}\left( {{h^*}{{\boldsymbol{x}}_{\text{c}}} + {{\boldsymbol{w}}_{\text{A}}}} \right) + {\rho _{\text{t}}}{\boldsymbol{t}}} \right) + {{\boldsymbol{w}}_{\text{E}}}} \right) + {{\boldsymbol{w}}_{\text{B}}}\notag\\
 =& {h_{{\text{EB}}}}{h_{{\text{AE}}}}{\rho _{\text{s}}}{h^*}{{\boldsymbol{x}}_{\text{c}}} + {h_{{\text{EB}}}}{h_{{\text{AE}}}}{\rho _{\text{t}}}{\boldsymbol{t}}\notag\\
 & +{\underbrace{ {h_{{\text{EB}}}}{h_{{\text{AE}}}}{\rho _{\text{s}}}{{\boldsymbol{w}}_{\text{A}}} + {h_{{\text{EB}}}}{{\boldsymbol{w}}_{\text{E}}} + {{\boldsymbol{w}}_{\text{B}}}}_{\boldsymbol{w}_{\text{TBCR}|\text{Eve}}}},
\end{align}
where ${h_{{\text{EB}}}}$ and ${h_{{\text{AE}}}}$ are  the channel coefficients between Eve and Bob, and between Alice and Eve, respectively. When no replay attack occurs, the received signal at Bob is given by
\begin{align}
{{\boldsymbol{y}}_{{\text{TBCR|Non}}}} =& h\left( {{\rho _{\text{s}}}\left( {{h^*}{{\boldsymbol{x}}_{\text{c}}} + {{\boldsymbol{w}}_{\text{A}}}} \right) + {\rho _{\text{t}}}{\boldsymbol{t}}} \right) + {{\boldsymbol{w}}_{\text{B}}}\notag\\
 =& ||h|{|^2}{\rho _{\text{s}}}{{\bf{x}}_{\text{c}}} + h{\rho _{\text{t}}}{\boldsymbol{t}} + {\underbrace{h{\rho _{\text{s}}}{{\boldsymbol{w}}_{\text{A}}} + {{\boldsymbol{w}}_{\text{B}}}}_{\boldsymbol{w}_{\text{TBCR}|\text{Alice}}}}.
\end{align}
Note that the noise at Eve is included in ${\boldsymbol{w}_{\text{TBCR}|\text{Eve}}}$. This results in a notable discrepancy between the noise distributions of ${\boldsymbol{w}_{\text{TBCR}|\text{Eve}}}$ and ${\boldsymbol{w}_{\text{TBCR}|\text{Alice}}}$. Thus, the advanced detection methods proposed in \cite{AgainstReply} can be employed to detect potential replay attacks. Similarly, the SCA scheme can employ the same approach for detecting replay attacks, which will not be elaborated further here.
\vspace{-0.5cm}
\subsection{Further Discussion}
In this subsection, we provide a further analysis of the proposed schemes in terms of RBE, time complexity, resource consumption, and applicability.

\begin{table}[t]
	\centering
	\caption{Comparison of time complexity between the proposed schemes and the typical SUP scheme.}
	\label{table 1}
	\vspace*{0.2cm}
	\begin{threeparttable}
		\setlength{\tabcolsep}{-0.2cm} 
		\begin{tabular}{ 
				>{\centering\arraybackslash}p{0.1\textwidth} 
				>{\centering\arraybackslash}p{0.25\textwidth}
				>{\centering\arraybackslash}p{0.2\textwidth} 
			}
			\toprule[1.5pt]
			\textbf{Scheme} & 
			\textbf{Source} & 
			\makecell{\textbf{Time Complexity}\textbf{(Tx/Rx)}} \\
			\hline
			SUP & 
			Decoding\tnote{1} and Test Statistic & 
			$\mathcal{O}(L\text{log}_2L)$/$\mathcal{O}(NL\text{log}_2L)$ \\
			\hline
			TBCR & 
			Test Statistic & 
			$\mathcal{O}(L)$/$\mathcal{O}(L)$ \\
			\hline
			SCA & 
			Folding Strategy and Test Statistic & 
			$\mathcal{O}(L)$/$\mathcal{O}(L)$ \\
			\bottomrule[1.5pt]
		\end{tabular}
		\begin{tablenotes}
			\footnotesize
			\item [1] We choose polar code for its low decoding complexity and the Successive Cancellation List (SCL) decoding algorithm is adopted\cite{SCLPOLAR}.
		\end{tablenotes}
	\end{threeparttable}
	\vspace*{-0.4cm}
\end{table}
\subsubsection{Ratio of Bandwidth Efficiency}
We compare the proposed schemes with the SUP scheme in terms of the RBE, which is defined as the ratio of the bandwidth efficiency with tagged signal $\eta_{\text{avg}}$ to that without tag embedding $\eta_{\text{0}}$, i.e.,
\begin{align}
\xi  \overset{\triangle}{=} \frac{{\eta_{{\text{avg}}}}}{{\eta_0}}=\frac{C_\text{avg}}{C_0},
\end{align}
where $C$ denotes the channel capacity. 
For the SUP scheme, the SNR with and without the embedded tag can be given by ${\Gamma _{{\text{Tag}}}} = {\rho _\text{s}^2}/\left({\sigma _{\text{B}}^2 + \rho _\text{t}^2}\right)$ and ${\Gamma _{{\text{0}}}} = 1/{\sigma _{\text{B}}^2}$, respectively. To prevent detection by Eve, every $N$ frames contains one authentication signal. Since the SUP scheme superimposes the tag onto the message, the length of the tagged signal is $L_2$, while that of the signal without the superimposed tag is $N\left(L_1+L_3\right)-L_2$. Thus, the average rate is expressed as
\begin{align}
{C_{{\text{avg,SUP}}}} = \frac{{\left( {N\left( {{L_1} + {L_3}} \right) - {L_2}} \right)\times C_{{\text{0}}} + {L_2}C_{{\text{Tag}}}}}{{N\left( {{L_1} + {L_3}} \right)}},
\end{align}
where $C_{{\text{0}}}=B\text{log}_2\left(1+\Gamma_{{\text{0}}}\right)$, $ C_{{\text{Tag}}}=B\text{log}_2\left(1+\Gamma_{{\text{Tag}}}\right)$, and $B$ is the bandwidth. $L_1$, $L_2$, and $L_3$ are the length of the pilot, authentication, and message signals, respectively. Thus, the RBE of the SUP scheme is expressed as
\begin{align}
{\xi _{{\text{SUP}}}} = \frac{{{C_{{\text{avg,SUP}}}}}}{C_0} = 1 - \frac{{\left( {1 - \Delta } \right){L_2}}}{{N\left( {{L_1} + {L_3}} \right)}},
\end{align}
where $\Delta  = C_{\text{Tag}}/C_0<1$.

For the TBCR scheme, the length of the signal without the superimposed tag is $N\left(L_1+L_3\right)+L_2$ ($L_2$ for the challenge and $N\left(L_1+L_3\right)$ for the response). In the response, the authentication signal with length $L_2$ contains no new information since it is a forwarded challenge signal. Thus, the average rate can be expressed as
\begin{align}
{C_{{\text{avg,TBCR}}}} = \frac{{\left( {N\left( {{L_1} + {L_3}} \right) + {L_2}} \right)\times C_{{\text{0}}} }}{{N\left( {{L_1} + {L_3}} \right)+2L_2}},	
\end{align}
and the RBE of the TBCR scheme can be expressed as
\begin{align}
{\xi _{{\text{TBCR}}}} = 1 - \frac{{{L_2}}}{{N\left( {{L_1} + {L_3}} \right) + 2{L_2}}}.
\end{align}

For SCA scheme, the effective signal length (pilot and message) is $N\left(L_1+L_3\right)$, where the length of the tagged signal is $L_2$. Thus, the average rate is expressed as
\begin{align}
  {C_{{\text{avg,SCA}}}} = \frac{{\left( {N\left( {{L_1} + {L_3}} \right) - {L_2}} \right)\times C_{{\text{0}}}+L_2\times C_\text{Tag} }}{{N\left( {{L_1} + {L_3}} \right)+L_2}},
\end{align}
and the RBE of the SCA scheme is given by 
\begin{align}
{\xi _{{\text{SCA}}}} = 1 - \frac{{(2 - \Delta ){L_2}}}{{N\left( {{L_1} + {L_3}} \right) + {L_2}}}.
\end{align}

\begin{remark}
The ratios of the RBEs of the proposed schemes to that of the SUP scheme are given by
\begin{align}
\!{R_{{\text{TBCR}}}}\! \!=\!\! \frac{{{{ {N^2\left( {{L_1} + {L_3}} \right)\left( {{L_1} + {L_3}+L_2} \right)} }}}}{{\left( {N\left( {{L_1}\! +\! {L_3}} \right)\! +\! 2{L_2}} \right)\!\left( {N\left( {{L_1}\!\! +\!\! {L_3}} \right)\!\! -\! \!\left( {1\! -\! \Delta } \right){L_2}} \right)}},
\end{align}
and
\begin{align}
{R_{{\text{SCA}}}} = \frac{{N\left( {{L_1} + {L_3}} \right)}}{{N\left( {{L_1} + {L_3}} \right) + {L_2}}}.
\end{align}
For example, when we set $N=10$ as the typical parameter in the 3GPP protocol\cite{3gpp}, we have $R_{\text{TBCR}}\approx 97.22\%$ and $R_{\text{SCA}}\approx 96.77\%$ with the tag power $\rho_\text{t}^2=0.1$, SNR $=0$ dB, $L_1=L_2=32$, and $L_3=64$. Moreover, both $R_{\text{TBCR}}$ and $R_{\text{SCA}}$ tend to 1 as $N$ increases. This indicates that our proposed schemes significantly improve detection performance at the cost of only a slight reduction in RBE.	
\end{remark}
\vspace*{-0pt}
\subsubsection{Time Complexity}\label{TC}

In this subsection, we compare the time complexity of the proposed schemes with that of the SUP scheme. At the transmitter, the SUP scheme superimposes the tag onto the encoded signal. Thus, its complexity mainly depends on the coding algorithm. In contrast, the complexity of the proposed schemes is linear, e.g., for interleaving, serial-to-parallel conversion. At the receiver, the SUP scheme, which estimates the tag through decoding, primarily incurs complexity from the decoding and test statistics. In contrast, the two proposed schemes exhibit reduced time complexity as the decoding is no longer needed. Specifically, the complexity of the TBCR scheme stems from the test statistics while that of the SCA scheme is mainly attributed to the folding strategy and the test statistics. Due to the linear complexity of both the de-interleaving and the folding strategy, the complexity of the test statistics in the SCA scheme remains $\mathcal{O}(L)$. The time complexity of different schemes is shown in TABLE \ref{table 1}, where $L$ and $N$ are the length of the authentication signal and the size of the decoding list, respectively. We can observe that the time complexity of the proposed schemes is linear, which is significantly lower than that of the SUP scheme.

\subsubsection{Consumption on Delay, Energy, and Keys}\label{DEK}
To conduct a more in-depth investigation of the proposed schemes, the authentication delay and energy consumption are analyzed.

First, we analyze the authentication delay. In the SUP scheme, the transmission delay, which includes $L_1$ pilot symbols and $L_3$ message symbols, and the propagation delay are denoted as $z_{\text{s}}$ and $z_{\text{p}}$, respectively.  To evade Eve’s detection, every $N$ frames contains one authentication signal. Thus, the authentication delay of the SUP scheme is expressed as
\begin{align}
	Z_\text{SUP}=N\left(z_{\text{s}}+z_{\text{p}}\right),
\end{align}	
where the processing delay defined by the time complexity is analyzed separately in \ref{TC} and is not included. Compared to the SUP scheme, the TBCR scheme consumes additional communication rounds in the challenge phase and transmit extra $L_2$ symbols in the response. Thus, the authentication delay of the TBCR scheme can be given by
\begin{align}
Z_{{\text{TBCR}}} = \frac{{ N\left({L_1} + {L_3}\right)+ 2{L_2}  }}{{{L_1} + {L_3}}}{z_{{\text{s}}}} + \left(N+1\right){z_{{\text{p}}}}.
\end{align}
The SCA scheme transmits extra $L_2$ symbols in the series signal and the authentication delay can be expressed as 
\begin{align}
	Z_{{\text{SCA}}} = \frac{N\left( L_1 + L_3\right) + L_2}{{{L_1} + {L_3}}}{z_{{\text{s}}}} + N{z_{{\text{p}}}}.
\end{align}

Then, we analyze the energy consumption. The energy for $N$ frames contained one authentication signal in the SUP scheme is denoted as $P_{\text{SUP}}=NE$, where $E$ denotes the energy consumed by one frame ($L_1$ pilot and $L_3$ message symbols included). Thus, the energy consumption of the TBCR and SCA schemes can be respectively given by
\begin{align}
{P_{{\text{TBCR}}}} = \frac{{N\left( {L_1} + {L_3} \right)+ 2{L_2} }}{{{L_1} + {L_3}}}E,
\end{align}
and
\begin{align}
{P_{{\text{SCA}}}} = \frac{{N\left( {L_1} + {L_3}\right) + {L_2} }}{{{L_1} + {L_3}}}E.
\end{align}
For example, when $L_1=L_2=32$, $L_3=64$, and $N=10$ \cite{3gpp}, we have $Z_\text{SUP}=10\left(z_\text{s}+z_\text{p}\right)$ while $Z_\text{TBCR}\approx 10.67z_\text{s}+11z_\text{p}$ and $Z_\text{SCA}\approx 10.33z_\text{s}+10z_\text{p}$. Moreover, the ratios of the energy are ${P_{{\text{TBCR}}}}/ P_{\text{SUP}}\approx 1.067$ and ${P_{{\text{SCA}}}}/ P_{\text{SUP}}\approx 1.033$. In practical scenarios, the TBCR scheme can be combined with ACK signals, which is discussed in the next subsection, and the length of the authentication signal in the SCA scheme can be flexibly shortened as needed. Thus, the proposed schemes significantly enhance authentication performance while consuming only a small amount of additional energy and time.

Finally, since the SCA scheme requires an interleaver, the key update frequency of the SCA scheme is twice that of the SUP and TBCR schemes.


\subsubsection{Applicability}


In this subsection, we discuss the applicability of the proposed schemes. The proposed schemes demonstrate superior performance in low SNR regions. In contrast, typical SUP scheme remains more effective in high SNR regions. Given the authentication performance of the proposed schemes, they are well-suited for networks that require high authentication performance, such as industrial control networks\cite{Industry} and military communication\cite{mil}, where a single false authentication may result in immeasurable consequences. In terms of compatibility, the proposed schemes can be integrated with existing communication systems. For example, the TBCR scheme can leverage the common ACK/NACK signals in communication systems \cite{ACK}, where Alice uses Bob’s ACK/NACK as the challenge to embed the tag. Similarly, the SCA scheme can integrate with spread-spectrum systems \cite{SS} by embedding the tag on two identical spreading codes. Given the broader context and focus of this paper, a detailed exploration is not pursued here.

\begin{remark}
	Considering authentication performance, RBE, time complexity, and additional resource consumption, the proposed schemes significantly enhance authentication performance and reduce time complexity while sacrificing only a small amount of RBE and additional resource. (cf. \textit{Observation \ref{OB2}} in Section \ref{5})
\end{remark}

\vspace*{-0.3cm}
\section{Numerical Results}\label{5}

In this section, we first provide an overview of the numerical simulation, including the parameter setup. Then, the robustness and security of the proposed TBCR and SCA schemes are simulated and analyzed respectively. Finally, we highlight the advantages of the proposed schemes and outline the specific application scenarios for both schemes, facilitating choosing the most suitable scheme based on the needs.

\vspace*{-0.3cm}

\subsection{Experimental Setup}
In this subsection, the proposed TBCR and SCA schemes are comprehensively compared with the typical SUP, BTP \cite{BTP}, CRAM \cite{CRAM} and BSUP \cite{XieBlind} schemes in the block fading channel. In the SUP and BTP schemes, decoding is employed to reconstruct the tag. The CRAM scheme employs the CR mechanism for authentication, while the BSUP scheme authenticates by superimposing the tag on the pilot. Specifically, the following parts are included in our numerical results. First, the performance degradation in detection probability between the ideal results (error-free decoding is assumed) and the actual performance (decoding errors lead to message interference) of the SUP scheme is simulated, which provided the motivation for our proposal of the TBCR and SCA schemes. Second, the theoretical results and simulations of the TBCR and SCA schemes are compared to verify the correctness of our theoretical derivations. Third, the comparisons of the performance between the TBCR scheme, the SCA scheme, the SUP scheme, and the BTP scheme are simulated in terms of robustness and security by the detection probability, Receiver Operating Characteristic (ROC) curve, and key equivocation. Fourth, we also explore the robustness of the proposed schemes against the channel estimation and synchronization errors,  and simulate the RBE to conduct a comprehensive evaluation of the proposed schemes.
Unless otherwise specified, we set the signal length $L=64$ and tag power $\rho_\text{t}^2=0.1$. Polar code is adopted \cite{PolarCode} for the simulations of the actual performance and each simulation is achieved by $M = 10^5$ times of independent Monte Carlo experiments. 

\vspace*{-0.5cm}
\subsection{Performance Degradation Evaluation}

\begin{figure}[!tbp]
	\setlength{\abovecaptionskip}{0pt}
	\centering
	\captionsetup{font=scriptsize}
	\includegraphics [width=3.5in]{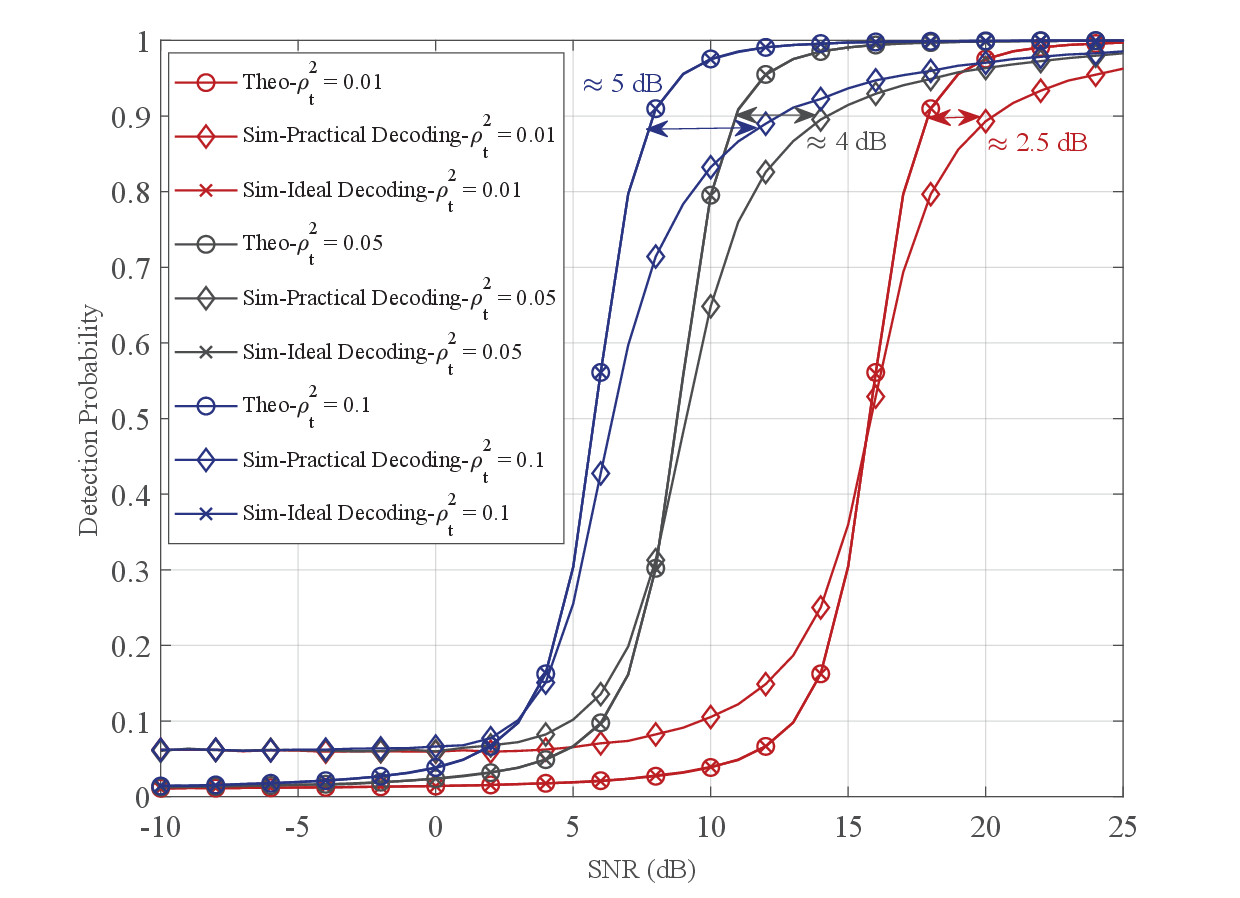}
	\caption{The performance gap between the theoretical results and simulations (considering the decoding errors or not) of the SUP scheme under different SNRs, where the performance degradation of the simulations (considering the decoding errors) becomes more pronounced as the tag power increases.}
	\label{PerformanceGAPdecoding}
	\vspace*{-0.5cm}
\end{figure}

\begin{observation}
\textit{In the typical SUP scheme, there is a performance gap between the actual performance, which accounts for decoding errors, and the ideal performance, which neglects these errors.  As the tag power increases, this performance gap widens. Thus, decoding errors have a significant impact on the performance of the typical SUP scheme, leading to sub-optimal robustness in practical applications. (cf. Fig. \ref{PerformanceGAPdecoding})}
\end{observation}

 The degradation between actual performance and ideal analysis in the tag-based PLA schemes is investigated under different tag powers. The numerical results are shown in Fig. \ref{PerformanceGAPdecoding}, where the polar code with a code length of 64 and a code rate of 1/2 is utilized. The threshold and theoretical performance of the SUP scheme are given by (41) and (44) in \cite{XiePSA}. From Fig. \ref{PerformanceGAPdecoding}, the following phenomena can be observed. First, the theoretical results match the simulations perfectly when error-free decoding (ideal decoding) is assumed. However, there is a gap between the theoretical results and the simulations when the decoding errors are considered (practical decoding). Second, the greater the tag power, the greater the performance gap. For example, with a 90\% detection probability as the baseline, the performance gap is about 2.5 dB when the tag power $\rho_\text{t}^2$ is set to 0.01. The performance gap is increased to 5 dB approximately when the tag power $\rho_\text{t}^2$ is set to 0.1. The reason for the performance gap is as follows. In tag-based PLA schemes, the tag is estimated by decoding the message and then removing the estimation of the message from the received signal. Compared with the normal signal ($\rho_\text{t}^2=0$), the non-zero tag power degrades the SNR, leading to imperfect decoding, which potentially introduces message interference and then the inaccurate estimation of the tag. Moreover, the performance gap widens as the tag power increases.

 Thus, in the SUP scheme, the message interference introduced by decoding errors degrades the detection performance, and the threshold derived is sub-optimal. The TBCR and SCA schemes are proposed to solve these limitations.
 
   \begin{figure}[!t]
 	\setlength{\abovecaptionskip}{0pt}
 	\centering
 	\captionsetup{font=scriptsize}
 	\includegraphics [width=3.5in]{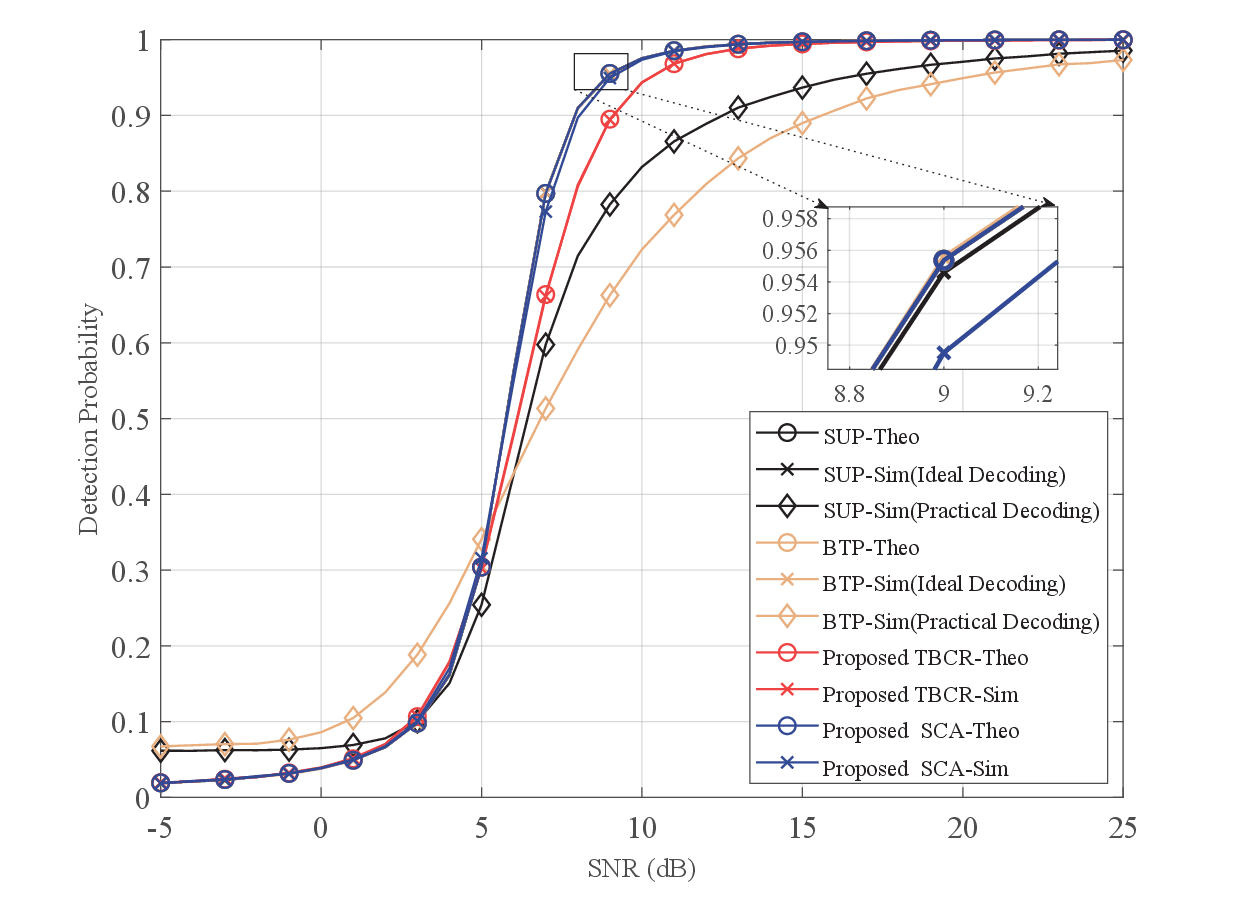}
 	\caption{Performance comparison of various schemes in terms of the detection probability versus the SNR at Bob, where $L=64$, $\rho_\text{t}^2=0.1$, and the SNR at Alice is 5 dB.}
 	\label{PDSNR}
 	\vspace*{-0.15cm}
 \end{figure}

\vspace*{-0.5cm}
\subsection{Robustness Evaluation}
\begin{figure}[!t]  
 	\centering
 	\captionsetup{font=scriptsize}
 	\subfigcapskip=0pt
 	\subfigure[SNR = 5 dB]{
 		\begin{minipage}[t]{0.24\textwidth}
 			\includegraphics[width=1\textwidth]{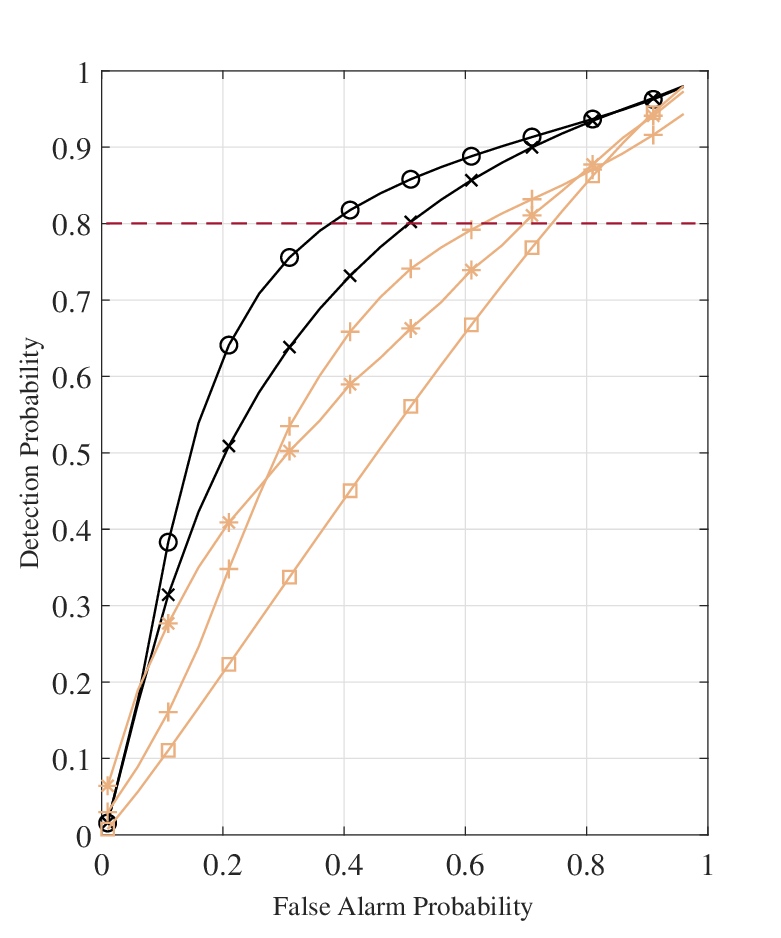} \\	
 		\end{minipage}
 		\label{ROC5}
 	}
 	\hspace{-7mm}
 	\subfigure[SNR = 7 dB]{
 		\begin{minipage}[t]{0.24\textwidth}
 			\includegraphics[width=1\textwidth]{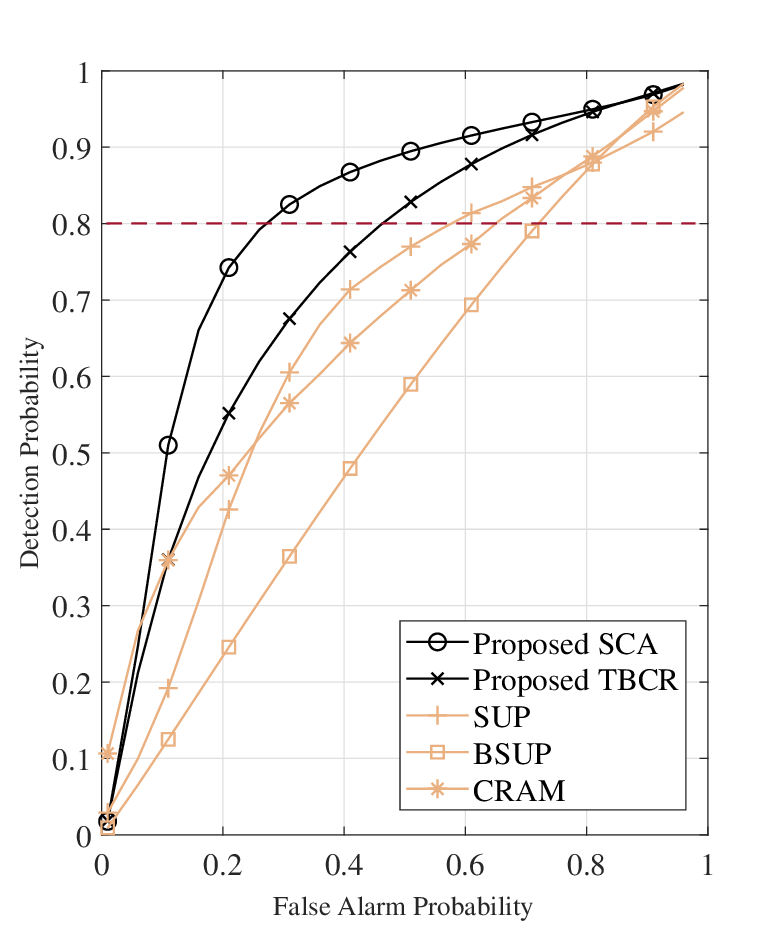} \\
 		\end{minipage}
 		\label{ROC7}
 	}
 	\caption{Comparison of ROC curve for the different schemes under (a) SNR = 5 dB and (b) SNR = 7 dB, where $L=64$, $\rho_\text{t}^2=0.01$, and the SNR at Alice is 10 dB.}
 	\label{ROC}
 	\vspace{-0.3cm}
 \end{figure}

\begin{figure}[!t]
 	\setlength{\abovecaptionskip}{0pt}
 	\centering
 	\captionsetup{font=scriptsize}
 	\includegraphics [width=3.5in]{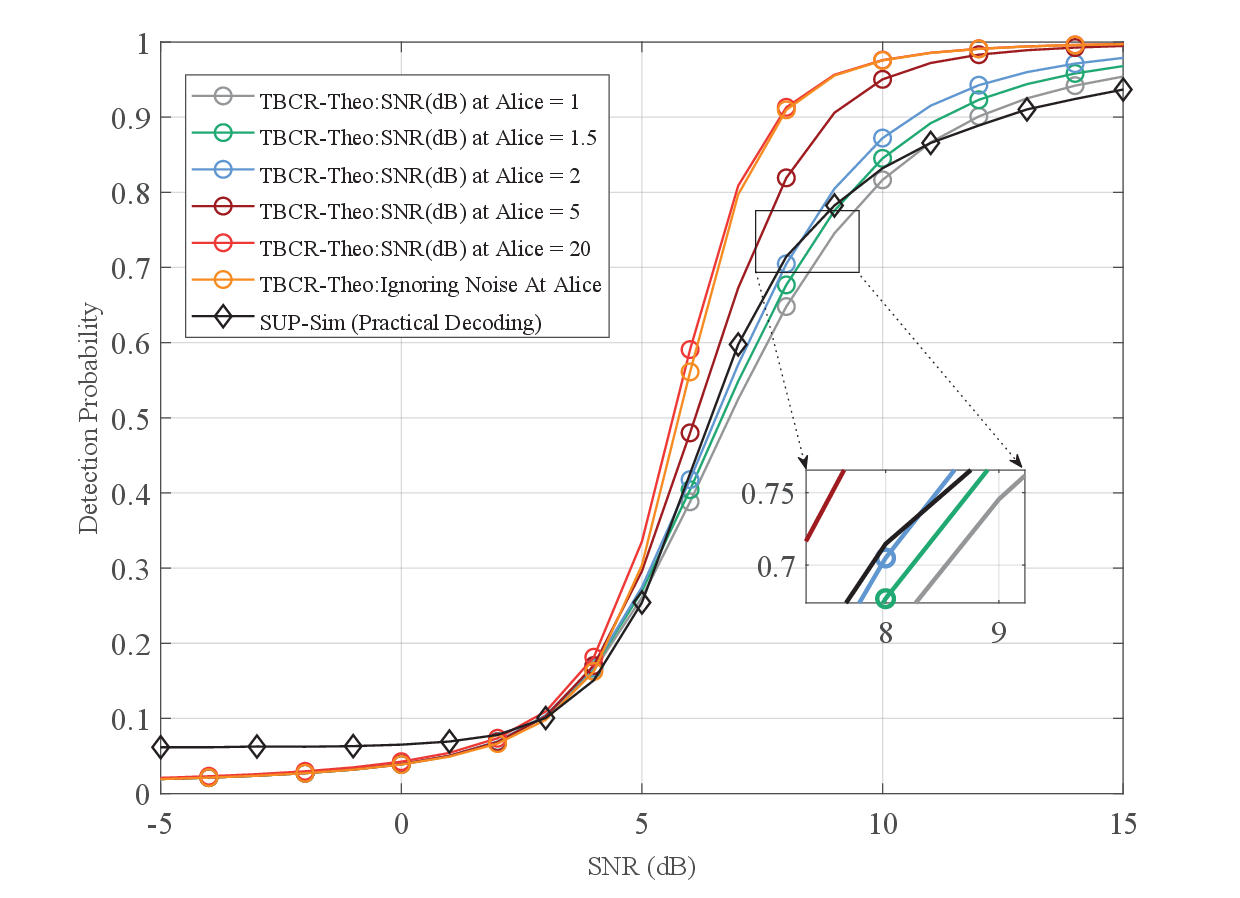}
 	\caption{Comparison of detection probability at Bob under varying SNRs at Alice for the TBCR scheme, where $\rho_\text{t}^2=0.1$ ,$P_{\text{FA}}=0.01$ and $L=64$.}
 	\label{TBCRAliceNoise}
 	\vspace{-0.1cm}
 \end{figure}

\begin{observation}\label{OB2}
\textit{Compared with the SUP and BTP schemes (considering decoding errors), the TBCR and SCA schemes demonstrate stronger robustness.    Notably, the SCA scheme achieves the ideal performance of the SUP scheme (neglecting decoding errors). Meanwhile, the performance of the TBCR scheme improves with increasing SNRs at Alice, ultimately approaching the ideal performance of the SUP scheme. (cf. Figs. \ref{PDSNR}, \ref{ROC} and \ref{TBCRAliceNoise})}
\end{observation}

 As shown in Fig. \ref{PDSNR}, the detection performance of different schemes is investigated, where we set $\rho_\text{t}^2=0.1$, $P_{\text{FA}}=0.01$ and the SNR at Alice is 5 dB. The theoretical results of the TBCR and SCA schemes are given by (\ref{eqTBCR_PD}) and (\ref{PD_SCA}), respectively. The theoretical results of the BTP scheme are given by (14) in \cite{BTP}. Our observations are as follows. First, the theoretical performance results of the TBCR and SCA schemes match perfectly with the simulations, confirming the accuracy of the derivations. Second, the TBCR scheme outperforms the simulation performance of the SUP and BTP schemes when the decoding errors are considered (practical decoding), and the SCA scheme achieves the ideal performance (ideal decoding) of the typical schemes. The reason for this phenomenon is that the TBCR and SCA schemes avoid decoding by carefully designing the authentication process and the signal frame, respectively. Thus, there is no performance degradation caused by decoding errors in the proposed schemes. However, the noise at Alice is introduced in the TBCR scheme and the receiver SNR at Bob is degraded. Thus, the performance of the TBCR scheme decreases slightly compared with the ideal performance of the typical schemes. \footnote{In scenarios with poor channel conditions (SNR < 0 dB), the standalone enhancement technique in \cite{CRH} can be combined with our proposed schemes.}
 
  We simulate the ROC curves of different schemes under varying SNRs, with the signal length $L=64$ and tag power $\rho_\text{t}^2=0.01$. To ensure fairness, we normalize the tag power in different schemes. From Fig. \ref{ROC}, the following phenomena can be observed. First, under the same SNR, the proposed SCA scheme exhibits the largest area under the ROC curve, indicating superior detection performance. The area under the ROC curve for the TBCR scheme is slightly lower than that of the SCA scheme due to the noise at Alice. The detection performance of the other schemes is inferior to that of the proposed schemes. Second, the area under the ROC curve for all schemes increases as the SNR increases.
  
  The detection performance of the TBCR scheme under different SNRs at Alice is simulated, where the noise at Alice is neglected corresponds to \textbf{Proposition \ref{pro1}}. As illustrated in Fig. \ref{TBCRAliceNoise}, the following phenomena can be observed. First, the higher the SNR at Alice, the better the detection probability. Second, when the SNR at Alice is about 20 dB, the detection probability achieves the ideal performance. Third, when the SNR at Alice is higher than 2 dB, the proposed TBCR scheme outperforms the typical SUP scheme.

\begin{figure}[!t]  
	\centering
	\captionsetup{font=scriptsize}
	\subfigcapskip=0pt
	\subfigure[Different signal length $L$]{
		\begin{minipage}[t]{0.24\textwidth}
			\includegraphics[width=1\textwidth]{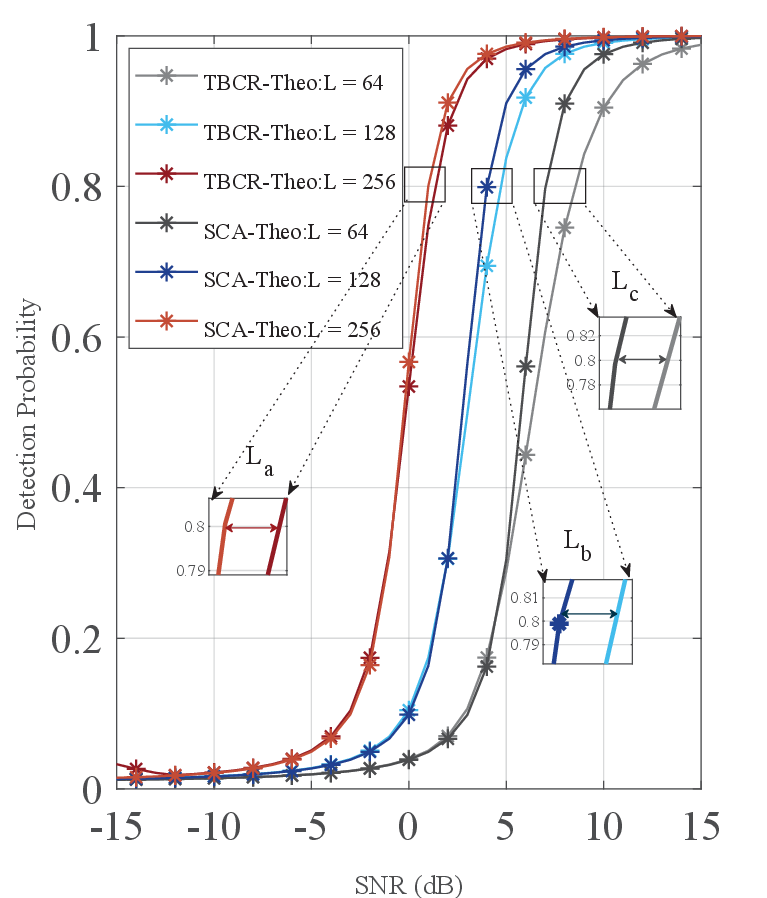} \\	
		\end{minipage}
		\label{subPD-L}
	}
	\hspace{-7mm}
	\subfigure[Different tag power $\rho_\text{t}^2$]{
		\begin{minipage}[t]{0.24\textwidth}
			\includegraphics[width=1\textwidth]{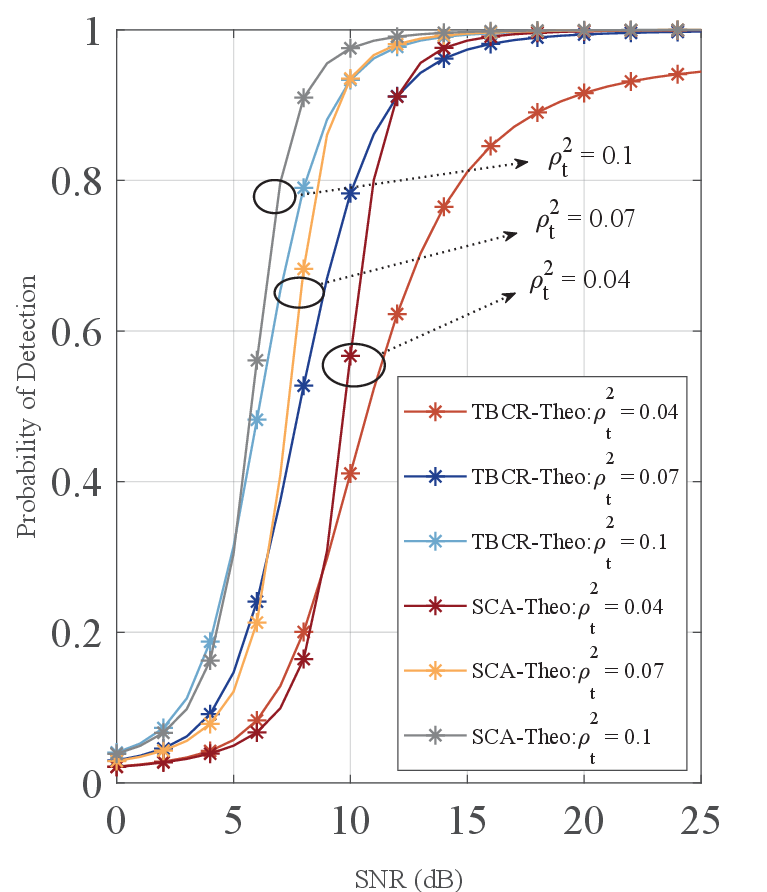} \\
		\end{minipage}
		\label{subPD-rhot}
	}
	\caption{Comparison of detection probability for the TBCR and SCA schemes under different (a) signal length $L$ and (b) tag power $\rho_\text{t}^2$.}
	\label{PDLrhot}
	\vspace{-0.2cm}
\end{figure}

\begin{figure}[!t]
	\setlength{\abovecaptionskip}{0pt}
	\centering
	\captionsetup{font=scriptsize}
	\includegraphics [width=3.5in]{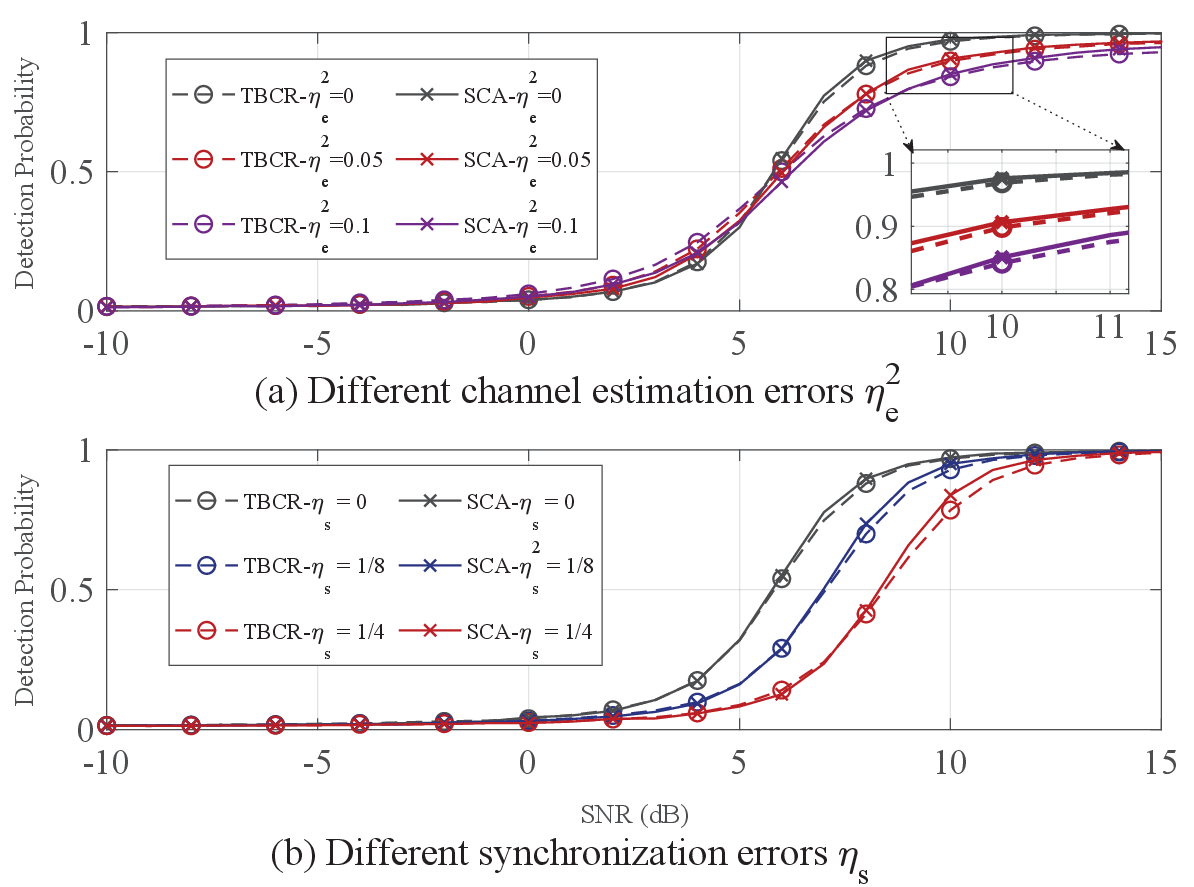}
	\caption{Comparison of detection probability for the TBCR and SCA schemes under different (a) channel estimation errors $\eta_\text{e}^2$ and (b) synchronization errors $\eta_\text{s}$.}
	\label{CSIIM}
	\vspace{-0.1cm}
\end{figure}

\begin{observation}
\textit{As the length of the signal and tag power increase, the detection probability of both the TBCR and SCA schemes improves, and the performance gap between the two schemes gradually narrows. Moreover, both the proposed schemes demonstrate strong robustness to channel estimation and synchronization errors. (cf. Figs. \ref{PDLrhot} and \ref{CSIIM})}
\end{observation}

The comparisons of the detection probability for the TBCR and SCA schemes under different signal lengths and tag power are illustrated in Fig. \ref{PDLrhot}, where we set $P_{\text{FA}}=0.01$ and the SNR at Alice is 3 dB. From Fig. \ref{subPD-L}, the following phenomena can be observed. First, the detection probability of both schemes improves by approximately 3 dB with a doubling of the signal length. Second, the SCA scheme outperforms the TBCR scheme due to the noise accumulation in the TBCR, and the gap between these two schemes decreases when the signal length increases. For example, when we set the 80\% detection probability as the baseline, the relation of the gap under different signal lengths is $L_\text{a}<L_\text{b}<L_\text{c}$. This phenomenon gives us the insight that the gap between the TBCR and SCA schemes can be reduced by increasing the signal length. Furthermore, the detection performance of the two schemes is investigated under different power allocations. It can be observed from Fig. \ref{subPD-rhot} that Bob's detection probability improves as the tag power increases for both the proposed schemes. However, eavesdroppers can more easily extract the tag from the intercepted signals with the increase of tag power. Thus, it is a topic worthy of exploration to balance security with the detection probability.

The effects of channel estimation errors and synchronization errors on the detection performance of the proposed schemes as simulated in Fig. \ref{CSIIM}, where we set $L=64$, $\rho_\text{t}^2=0.1$, and the SNR at Alice is 10 dB. In Fig. \ref{CSIIM}(a), the normalized CSI error $\eta_\text{e}^2$ indicates the extent of the channel estimation error. It can be observed that as the channel estimation error increases, the detection probability of both proposed schemes decreases. However, when $\eta_\text{e}^2<0.1$, the decrease in detection probability is relatively small for both schemes. In Fig. \ref{CSIIM}(b), the oversampling factor $N$ is set to 8, and the synchronization error indicates the offset relative to the symbol duration. It is evident that as the synchronization error increases, the detection probability of both proposed schemes decreases. However, when the synchronization error is less than 1/4, the detection probability of the proposed schemes experiences only a small reduction. In conclusion, the proposed schemes exhibit good robustness to both channel estimation errors and synchronization errors.
\vspace*{-15pt}
\subsection{Security Evaluation}
\vspace*{-5pt}
\begin{figure}[!tbp]
	\setlength{\abovecaptionskip}{0pt}
	\centering
	\vspace{-0.1cm}
	\captionsetup{font=scriptsize}
	\includegraphics [width=3.5in]{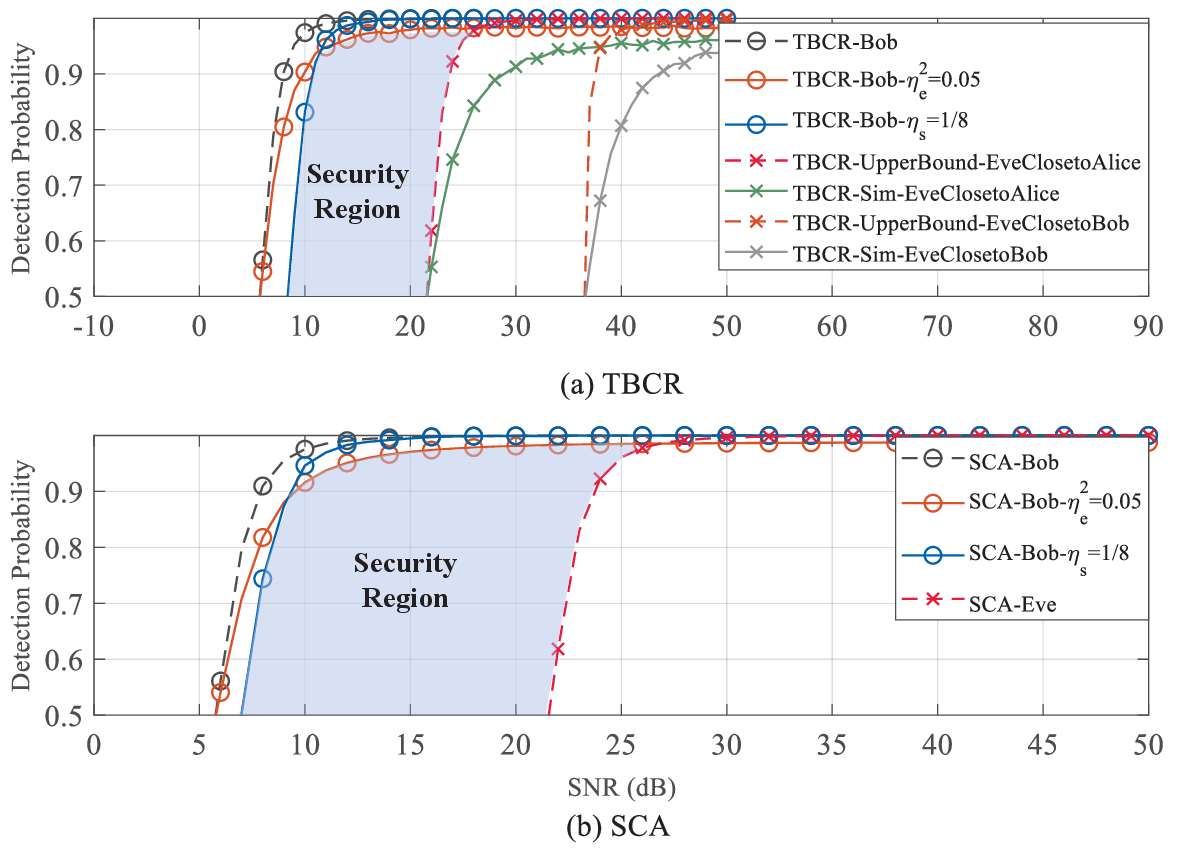}
	\caption{Comparison of detection probability at Bob and Eve under the (a) TBCR and (b) SCA schemes, where we set $L=64$ and $P_{\text{FA}}=0.01$, and the channel estimation and synchronization errors are considered. Moreover, in the TBCR scheme, a capable Eve who can eavesdrop on both the challenge and response signals is considered.}
	\label{TBCRSCA-BobEve}
	\vspace*{-2pt}
	\vspace*{-0.2cm}
\end{figure}

\begin{figure}[!t]
	\setlength{\abovecaptionskip}{0pt}
	\centering
	\captionsetup{font=scriptsize}
	\includegraphics [width=3.5in]{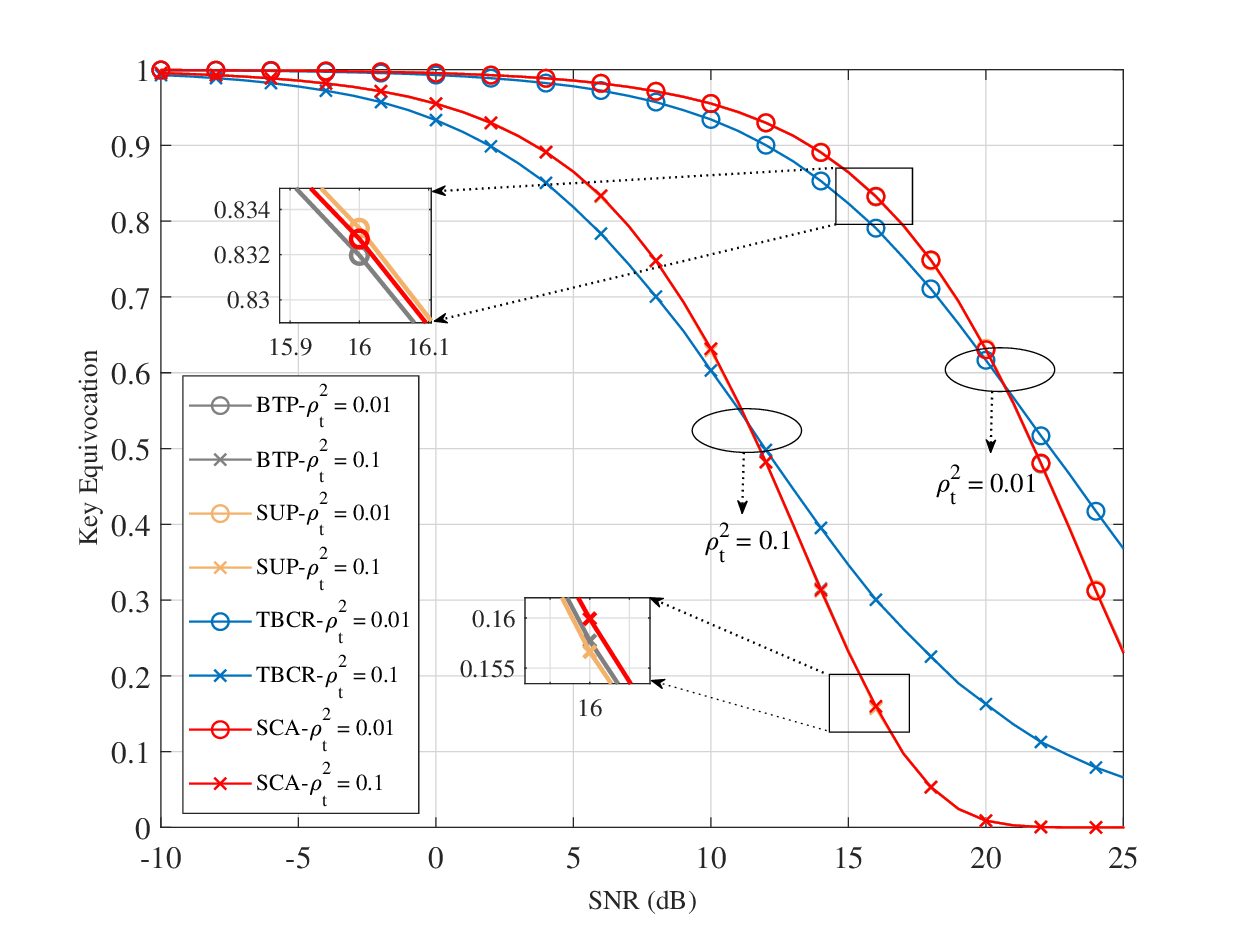}
	\caption{Comparison of the key equivocation of various schemes, where $L=64$, $P_{\text{FA}}=0.01$, and the SNR at Alice is set to 30 dB.}
	\label{KeyEquivocation}
	\vspace*{-0.3cm}
\end{figure}

\begin{observation}
\textit{In both the TBCR and SCA schemes, Eve's detection probability is significantly lower than Bob's, indicating strong security performance for both schemes. In terms of the key equivocation, the SCA scheme performs the same as the typical SUP scheme, while the TBCR scheme offers better security at high SNR regions. (cf. Figs. \ref{TBCRSCA-BobEve} and \ref{KeyEquivocation})}
\end{observation}


The difference in detection probability between Eve and Bob is compared in Fig. \ref{TBCRSCA-BobEve}, where it is assumed that the SNR at Eve is identical to that at Bob. We set the length of the signal $L=64$, tag power $\rho_\text{t}^2=0.1$, and the SNR at Alice is 10 dB. In the TBCR scheme, the theoretical results of the detection probability corresponding to Eve's close to Alice and Bob are given by (\ref{TBCR_EA}) and (\ref{TBCR_EB}), respectively. In the SCA scheme, the theoretical result of the detection probability at Eve is given by (\ref{SCA_E}). We can observe the following phenomena. First, for the TBCR scheme, the upper bound of Eve’s detection probability represents the optimal detection probability Eve can achieve, which is better than the practical performance. Second, when the practical conditions are considered, i.e., channel estimation and synchronization errors are present at Bob, his detection probability is lower than the performance without errors. Finally, even under practical conditions, Bob’s detection probability remains higher than Eve’s, with the difference being depicted by the ``Security Region”.

The variation of the key equivocation in different schemes according to SNR at Eve is investigated. The key equivocation is correlated with the tag power, hence we set $\rho_\text{t}^2=0.1$ and $\rho_\text{t}^2=0.01$, as shown in Fig. \ref{KeyEquivocation}. The theoretical results for the key equivocation of the TBCR and SCA schemes are given by (\ref{TBCRKE}). The following phenomena can be observed. First, for all the schemes, the lower the tag power, the higher the key equivocation. This is because the lower power tags are more conducive to hiding in the noise, thus evading Eve's detection. Second, the key equivocation of the proposed SCA scheme is comparable to that of the BTP and SUP schemes due to the identical TNR at Eve in these three schemes. For the TBCR scheme, its key equivocation is lower at low SNR regions but superior to the SCA scheme at high SNR regions. 
\vspace*{-15pt}
\subsection{Bandwidth Efficiency}
\vspace*{-5pt}

\begin{figure}[!t]
	\centering
	\captionsetup{font=scriptsize}
	\setlength{\subfigcapskip}{0pt}
	\setlength{\abovecaptionskip}{4pt}
	
	\subfigure[$N=1$]{
		\includegraphics[width=0.24\textwidth]{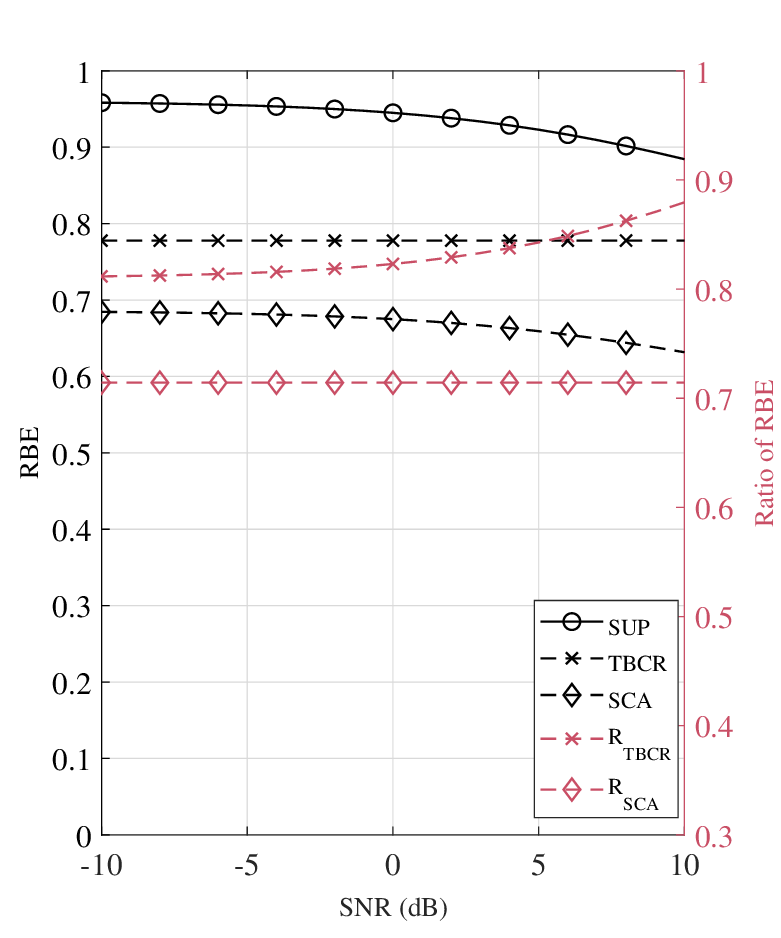}
		\label{RBEn1}
	}
	\hspace{-20pt}  
	\subfigure[$N=10$]{
		\includegraphics[width=0.24\textwidth]{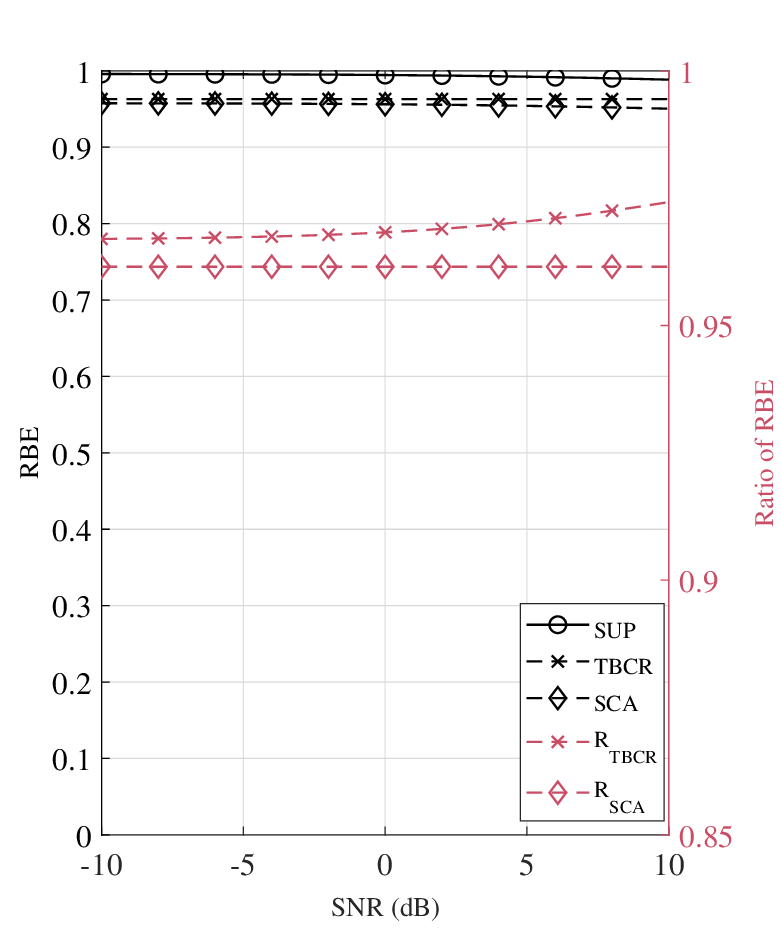}
		\label{RBEn10}
	}
	
	\caption{RBE of various schemes across the SNRs under (a) $N=1$ and (b) $N=10$. The left axis shows the RBE while the right axis shows RBE ratio to the SUP scheme.}
	\label{BE}
\end{figure}

We present a comparison of the RBE between the proposed schemes and the SUP scheme, with $\rho_t^2=0.1$, $L_1=32$, $L_2=64$, and $L_3=128$. From Fig. \ref{BE}, the following observations can be made. First, among the three schemes, the SUP scheme exhibits the highest RBE, while the SCA scheme has the lowest. Second, the RBE of the proposed schemes improves with increasing $N$. Third, although the proposed schemes incur some loss in RBE, the ratio of their RBEs to that of the SUP scheme remains above 70\%.  Thus, the proposed schemes maintain a higher detection probability with slight loss in RBE, and the RBE can be flexibly controlled by adjusting the ratio of authentication signal to message in practical deployments.
\vspace*{-0pt}

 \begin{figure*}[!t]
 	\begin{align}\nonumber\label{proofAind}
 		{P_{\text{FA},\text{TBCR}}} =& \mathbb{E}\left\{ {Q\left( {\frac{{\gamma \sqrt {2\rho_\text{t}^2|{h_k}{|^2}} }}{{\sqrt {L(\sigma_\text{A}^2|{h_k}{|^2} + \sigma_\text{B}^2)} }}} \right)} \right\}
 		=\int_0^\infty \! {\left(\! {\int_{\left(\! {\frac{{\gamma\! \sqrt {2\rho _{\text{t}}^2y} }}{{\sqrt {L\!(\sigma _{\text{A}}^2y\! +\! \sigma _{\text{B}}^2)} }}}\! \right)}^\infty \! {\frac{1}{{\sqrt {2\pi } }}{\text{exp}}\left(\! {\frac{{ - {x^2}}}{2}}\! \right)dx} } \right)} \frac{1}{{\sigma _{\text{h}}^2}}{\text{exp}}\left( {\frac{{ - y}}{{\sigma _{\text{h}}^2}}} \right)dy\notag\tag{A.6}\\
 		\overset{\text{A}}{=}&{\underbrace{\int_0^{\sqrt {\frac{a}{{\sigma _{\text{A}}^2}}} }\! {\left(\! {\int_0^{\frac{{\sigma _{\text{B}}^2{x^2}}}{{a - \sigma _{\text{A}}^2{x^2}}}}\!\! {\frac{1}{{\sigma _{\text{h}}^2}}{\text{exp}}\left( {\frac{{ - y}}{{\sigma _{\text{h}}^2}}} \right)\!dy} } \!\right)\!\!} \frac{1}{{\sqrt {2\pi } }}{\text{exp}}\!\left(\! {\frac{{ - {x^2}}}{2}} \!\right)\!dx \!}_{\text{I}_{1}}}+\! {\underbrace{\int_{\sqrt {\frac{a}{{\sigma _{\text{A}}^2}}} }^\infty \!\! {\left(\! {\int_0^\infty\!\!  {\frac{1}{{\sigma _{\text{h}}^2}}{\text{exp}}\left( {\frac{{ - y}}{{\sigma _{\text{h}}^2}}} \!\right)\!dy} }\! \right)\!} \!\frac{1}{{\sqrt {2\pi } }}{\text{exp}}\left( {\!\frac{{ - {x^2}}}{2}}\! \right)\!dx}_{\text{I}_{2}}}\notag\\
 		\overset{\text{B}}{=}& \left(\frac{1}{2} - Q\left( {\sqrt {\frac{a}{{\sigma _{\text{A}}^2}}} } \right) - {\underbrace{\int_0^{\sqrt {\frac{a}{{\sigma _{\text{A}}^2}}} } {\frac{1}{{\sqrt {2\pi } }}{\text{exp}}\left( { - \frac{{\sigma _{\text{B}}^2{x^2}}}{{\sigma _{\text{h}}^2\left( {a - \sigma _{\text{A}}^2{x^2}} \right)}} - \frac{{{x^2}}}{2}} \right)dx}}_\Omega} \right) + Q\left( {\sqrt {\frac{a}{{\sigma _{\text{A}}^2}}} } \right)\tag{A.7}\label{A7}
 	\end{align}
 	\hrulefill \vspace*{-10pt}
 \end{figure*} \leavevmode \\
 
\vspace*{-1.1cm}
\section{conclusion}\label{6}
In this paper, the novel TBCR and SCA schemes were first proposed to address the limitations of the message interference and sub-optimal detection probability in existing tag-based PLA schemes. 
Subsequently, the closed-form expressions for both proposed schemes in terms of robustness and security were provided. In particular, on one hand, optimal threshold and optimal detection probability were derived, indicating that the SCA scheme can achieve the ideal performance (error-free decoding) of the SUP scheme, while the TBCR scheme approaches the ideal performance as the SNR at Alice increases. On the other hand, in terms of security, the TBCR scheme provides enhanced security at high SNR regions while requiring fewer keys. The numerical results demonstrate the superior robustness and security of the proposed schemes. In conclusion, the proposed schemes offer new perspectives and enhance the detection performance for tag-based PLA by optimizing the authentication process and constructing authentication signals. However, the proposed schemes still have certain limitations, such as the additional RBE consumption. These issues will be the focus of our future work, where we aim to further optimize the schemes for better efficiency and performance.

\vspace*{-10pt}

\section*{Appendix A}\label{appA}
\renewcommand{\thefigure}{A\arabic{figure}}
\setcounter{figure}{0} 

\begin{figure}[!tbp]  
	\centering
	\captionsetup{font=scriptsize}
	\vspace{-0cm}
	\subfigcapskip=-5pt
	\subfigure[The integral interval of (A.6)]{
		\begin{minipage}[t]{0.237\textwidth}
			\includegraphics[width=1\textwidth]{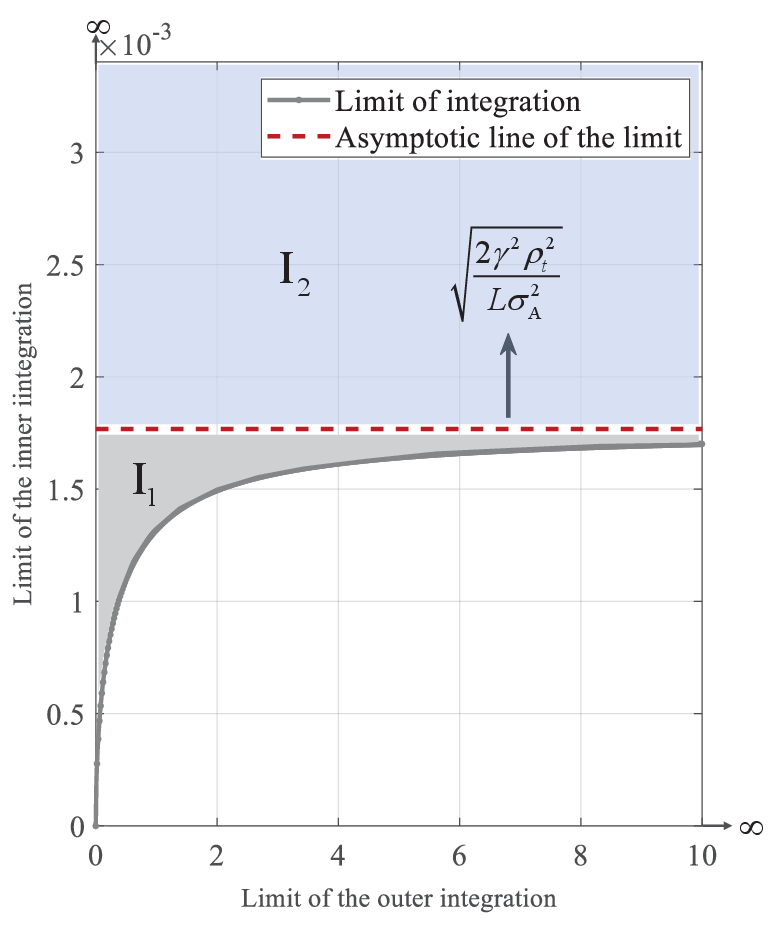} \\	
		\end{minipage}
		\label{Intergration_sub}
	}
	\hspace{-7mm}
	\subfigure[False alarm probability in (\ref{RWPFA})]{
		\begin{minipage}[t]{0.24\textwidth}
			\includegraphics[width=1\textwidth]{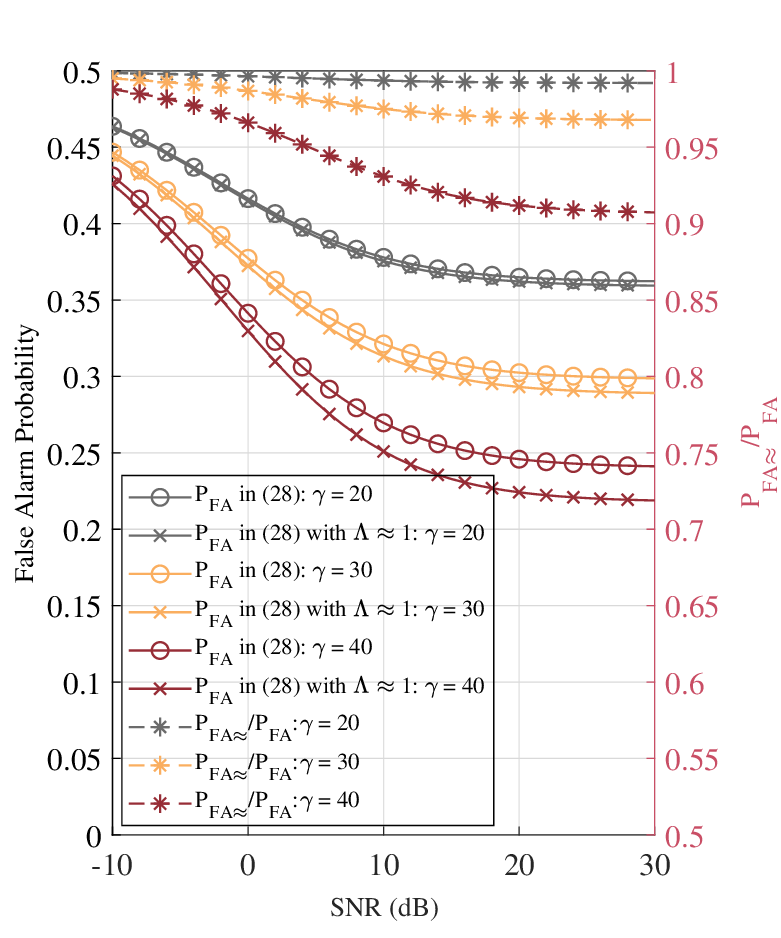} \\
		\end{minipage}
		\label{Proof_sub}
	}
	\caption{The figures used to assist the proof in APPENDIX A, including (a) The integral interval of (A.6), and (b) the comparison of the false alarm probability in (\ref{RWPFA}) and the approximated one. }
	\label{APPENDIXA_assit}
	\vspace{-0.2cm}
\end{figure}

The test statistic under different hypothesis can be obtained by substituting (\ref{r_TBCR}) into (\ref{eq1}) as
\begin{align}\nonumber
	{\tau _{\text{TBRC}|{\text{H}_0}}} = \frac{1}{{{\rho _\text{t}}}}{\left( {(1 - {\rho _\text{s}})h{\boldsymbol{x}_\text{c}} + {\boldsymbol{w}_{\text{A}}} + \frac{{{\boldsymbol{w}_{\text{B}}}}}{h}} \right)^{\dag}}\boldsymbol{t},
\tag{A.1}
\end{align}
and
\begin{align}\nonumber
	{\tau _{\text{TBCR}|{\text{H}_1}}} = {\left( {\boldsymbol{t} + \frac{{{\rho _\text{s}}{\boldsymbol{w}_{\text{A}}}}}{{{\rho_{\text{t}}}}} + \frac{{{\boldsymbol{w}_{\text{B}}}}}{{{\rho_{\text{t}}}h}}} \right)^{\dag}}\boldsymbol{t},
\tag{A.2}
\end{align}	
where $\rho _\text{s}$ and $\rho _\text{t}$ satisfy $\rho _\text{s}^2\sigma^2_{{\boldsymbol{y}_\text{A}}} + \rho _\text{t}^2 = 1$, and $\sigma _{{\boldsymbol{y}_\text{A}}}^2 = \sigma _\text{h}^2 + \sigma _\text{A}^2$ is the average power of $\boldsymbol{y}_{\text{A}}$. We assume that 
$\mathbb{E}(||{\boldsymbol{t}}{||^2}) = \mathbb{E}(||{\boldsymbol{x}_\text{c}}{||^2}) = L$. The receiver noise at Alice and at Bob, and the channel are modeled as $\boldsymbol{w}_{\text{A}}\sim \mathcal{CN}\left(0,\sigma_\text{A}^2\boldsymbol{I}_L\right)$, $\boldsymbol{w}_{\text{B}}\sim \mathcal{CN}\left(0,\sigma_\text{B}^2\boldsymbol{I}_L\right)$, and $h\sim \mathcal{CN}\left(0,\sigma_\text{h}^2\right)$, respectively. It is assumed that the challenge signal and the tags are independent of each other, i.e., $\mathbb{E}(\boldsymbol{x}_c^\dag\boldsymbol{t})=0$. Thus, we have
\begin{align}\nonumber\label{delta_TBCR_H0}
	{\delta _{\text{TBCR}|{\text{H}_0}}}\sim \mathcal{CN}\left(0,\frac{{L(\sigma _\text{A}^2|h{|^2} + \sigma_{\text{B}}^2)}}{{2\rho_\text{t}^2|h{|^2}}}\right),
\tag{A.3}
\end{align}
and
\begin{align}\nonumber\label{delta_TBCR_H1}
	{\delta _{\text{TBCR}|{\text{H}_1}}}\sim \mathcal{CN}\left(L,\frac{{L(\rho_\text{s}^2\sigma_\text{A}^2|h{|^2} + \sigma_\text{B}^2)}}{{2\rho_\text{t}^2|h{|^2}}}\right).
\tag{A.4}
\end{align}
Given the threshold $\gamma$ , the  $P_{\text{FA}}$ of Bob in the TBCR scheme for block $k$ can be expressed as
\begin{align}\nonumber\label{PFAi_Bob}
	{P_{\text{FA},\!\text{TBCR},k}}\! =\! {P_{{\text{H}_0}}}\!({\delta_{\text{TBCR}}}\!\! >\!\! \gamma )\! =\! Q\!\left(\!\!\frac{{\gamma \sqrt {2\rho_\text{t}^2|{h_k}{|^2}} }}{{\sqrt {L(\sigma_\text{A}^2|{h_k}{|^2}\! +\! \sigma_\text{B}^2)} }}\!\right)\!.
\tag{A.5}
\end{align}
Since the channel $h_k$ follows a CSCG distribution with zero-mean and variance $\sigma_\text{h}^2$, $|h_k|^2$ follows an exponential distribution with variance $\sigma_\text{h}^2$. Thus, the false alarm probability can be obtained by taking the expectation of (\ref{PFAi_Bob}) over the channel $h_k$ as (\ref{proofAind}), shown at the top of next page. For simplicity, let $a = 2{\gamma ^2}\rho _\text{t}^2/L$. The integral interval of (\ref{proofAind}), as illustrated in Fig. \ref{APPENDIXA_assit} (a), can be divided into two parts: $\text{I}_1$ and $\text{I}_2$. $\overset{\text{A}}{=}$ involves exchanging the order of integration in (\ref{proofAind}) and integrating separately over $\text{I}_1$ and $\text{I}_2$, and $\overset{\text{B}}{=}$ presents the corresponding results of the integration. Solving $\Omega$ in (\ref{A7}) presents a significant challenge due to $\sigma_{\text{A}}^2$. Thus, the Gaussian-Chebyshev Quadrature (GCQ) \cite{HAND} is applied to $\Omega$ to obtain (\ref{eqTBCR_PFA}), where the positive integer $N$ determines the precision. 

The traversal to (\ref{eqTBCR_PFA}) is recommended to obtain the threshold in the TBCR scheme. However, we propose a heuristic method to obtain the threshold at high SNRs (corresponding to a smaller threshold). We rewrite (\ref{eqTBCR_PFA}) as
\begin{align}\nonumber\label{RWPFA}
	{P_{{\text{FA,TBCR}}}} =& \frac{1}{2} - \frac{1}{2}\sqrt {\frac{a}{{\sigma _\text{A}^2}}} \frac{\pi }{N}\sum\limits_{i = 1}^N {\sqrt {1 - k^2(i)} }\notag\\
	&\times\! \frac{1}{{\sqrt {2\pi } }}\!\exp\! \left(\!\! { - \frac{{\sigma _\text{B}^2\left( {k^2(i) + 1 + 2k(i)} \right)}}{{4\sigma _\text{h}^2\sigma _\text{A}^2\left( {\frac{3}{4}\!\! -\!\! \frac{1}{4}k^2(i)\! -\! \frac{1}{2}k(i)}\! \right)}}}\!\! \right)\notag\\
	&\times{\underbrace{ \exp\left( { - \frac{a}{{8\sigma _\text{A}^2}}\left( {k^2(i) +1 + \frac{1}{2} k(i)} \right)} \right)}_\Lambda},
	\tag{A.8}
\end{align}
where $k(i)=\cos\frac{(2i-1)\pi}{2N}$ holds. Note that the power allocation $\rho_\text{t}^2$ is usually small in practical setups \cite{XieSurvey} and then $\Lambda \approx 1$ holds. Thus, the threshold can be obtained by separating $\gamma$ from (\ref{RWPFA}) with $\Lambda \approx 1$. To evaluate the validity of the $\Lambda \approx 1$ approximation, the comparison between the original formula (\ref{RWPFA}) and the approximated one $P_{\text{FA}\approx}$ is illustrated in Fig. \ref{APPENDIXA_assit} (b) by setting different $\gamma$. It can be seen from Fig. \ref{APPENDIXA_assit} (b) that the approximated PFA matches well with (\ref{RWPFA}) when we set $\gamma=20$, and there is a significant difference between (\ref{RWPFA}) and the approximated PFA when $\gamma=30$ and $\gamma=40$. Thus, the heuristic method is preferable with limited computational resources and a high SNR; otherwise, the traversal to (\ref{eqTBCR_PFA}) is recommended to find the optimal threshold.

Based on (\ref{delta_TBCR_H1}), the $P_\text{D}$ of Bob in the TBCR scheme for block $k$ can be expressed as
\begin{align}\nonumber\label{P_Di}
	{P_{\text{D},k}} = Q\left( {\frac{{({\gamma _{\text{TBCR}}} - L)\sqrt {2\rho_\text{t}^2|{h_k}{|^2}} }}{{\sqrt {L(\rho_\text{s}^2\sigma_\text{A}^2|{h_k}{|^2} + \sigma_\text{B}^2)} }}} \right).
\tag{A.9}
\end{align}
Similarly, the expectation of (\ref{P_Di}) over ${h}_k$ is taken to yield (\ref{eqTBCR_PD}). The proof of (\ref{TBCRNoAlice1}) to (\ref{TBCRNoAlice3}) is obtained by setting $\sigma_\text{A}^2=0$ in (\ref{delta_TBCR_H0}) and (\ref{delta_TBCR_H1}). 

The proof is completed.
 $\hfill\blacksquare$

\vspace*{-10pt}
\section*{Appendix B}\label{appB}
 According to (\ref{delta_SCA}), the test statistic under different hypotheses can be respectively expressed as 
\begin{align}\nonumber
	{\tau _{\text{SCA}|{\text{H}_0}}} =\frac{1}{2} \left({\frac{{{\boldsymbol{N}}_\text{B}}}{{{\rho _\text{t}}h}}}\right)^{\dag}{\boldsymbol{T}},
\tag{B.1}
\end{align}
\vspace{-0.1cm}
and	
\vspace{-0.1cm}
\begin{align}\nonumber
	{\tau _{\text{SCA}|{\text{H}_1}}} =\frac{1}{2} {\left({\boldsymbol{T}} + \frac{{\boldsymbol{N}_\text{B}}}{{{\rho _\text{t}}h}}\right)^{\dag}}{\boldsymbol{T}},
\tag{B.2}
\end{align}
where ${h}\triangleq{\hat{h}_{\text{BA}}} $. When BPSK is employed, the distribution of the random variable $\boldsymbol{T}_i$ $\left(i\in\{1, 2, 3, \cdots, L\}\right)$ is given as ${{P}}({\boldsymbol{T}_i} =  \pm 2) = 1/4$  and ${{P}}({\boldsymbol{T}_i} = 0) = 1/2$. Thus, we have $\mathbb{E}(\boldsymbol{T}_i)=0$  and $\text{var}(\boldsymbol{T}_i)=2$. Since $\boldsymbol{w}_{\text{B},i}\sim \mathcal{CN}(0,\sigma_\text{B}^2)$  and $\boldsymbol{w}_{\text{B},{i+L}}\sim \mathcal{CN}(0,\sigma_\text{B}^2)$  are independently and identically distributed (i.i.d.), it follows that $\boldsymbol{N}_{\text{B},i}\sim \mathcal{CN}(0,2\sigma _\text{B}^2)$. It is assumed that the noise and the tags are independent of each other, i.e., $\mathbb{E}(\boldsymbol{N}_\text{B}^{T}\boldsymbol{T})=0$. The probability distribution function of the test statistic in the SCA scheme can be modeled as
\begin{align}\nonumber
	{\delta _{\text{SCA}|{\text{H}_0}}}\sim \mathcal{CN}\left(0,\frac{L\sigma_\text{B}^2}{(2\rho_\text{t}^2|h{|^2})}\right),
\tag{B.3}
\end{align}
and
\begin{align}\nonumber\label{SCA_H_1}
	{\delta _{\text{SCA}|{\text{H}_1}}}\sim \mathcal{CN}\left(L,\frac{L\sigma_\text{B}^2}{(2\rho_\text{t}^2|h{|^2})}\right).
\tag{B.4}
\end{align}
The remainder of the proof is similar to the derivation of (\ref{eqTBCR_PFA}) and (\ref{eqTBCR_PD}) from (\ref{delta_TBCR_H0}) and (\ref{delta_TBCR_H1}) in Appendix A, and will not be reiterated here.

The proof is completed.
 $\hfill\blacksquare$
 
\vspace*{-12pt}
\section*{Appendix C}
Here, a capable Eve who can eavesdrop on both the challenge and response signals is considered.
When Eve is close to Alice, $h_{\text{EA}}\approx 1$ and $h_{\text{BE}}\approx h_{\text{BA}}\triangleq h$ are assumed as in \cite{CRAM}. 
Thus, the received signal at Eve can be expressed as
\begin{align}\nonumber
	{\boldsymbol{y}_{\text{E}|{\text{H}_0}}} \approx h^*{\boldsymbol{x}_\text{c}} + {\boldsymbol{w}_{\text{A}}} + {\boldsymbol{w}_{\text{E}}},
	\tag{C.1}
\end{align}
and
\begin{align}\nonumber
	{\boldsymbol{y}_{\text{E}|{\text{H}_1}}} = {\rho_\text{s}}(h^*{\boldsymbol{x}_\text{c}} + {\boldsymbol{w}_{\text{A}}}) + {\rho_\text{t}}\boldsymbol{t} + {\boldsymbol{w}_{\text{E}}},
	\tag{C.2}
\end{align}
where $\text{H}_0$ means that no authentication is taking place while $\text{H}_1$ means that authentication is currently taking place. $\boldsymbol{w}_{\text{E}}\sim \mathcal{CN}(0,\sigma_{\text{E}}^2)$ is the receiver noise at Eve.
Eve may detect the authentication process by detecting the challenge signal due to his lack of knowledge regarding the secret keys \cite{XiePSA}. The challenge signal is estimated as 
\begin{align}\nonumber
	{\hat{\boldsymbol{x}}_{\text{AE}}} = \frac{{h}}{{|{{h}}{|^2}}}{\boldsymbol{y}_\text{E}}.
	\tag{C.3}
\end{align}
Since Eve can eavesdrop on both the challenge signal $\boldsymbol{x}_\text{c}$ and response signal, the residual signal is obtained by removing the challenge signal from the received response signal, i.e., ${\boldsymbol{r}_\text{E}} = {\boldsymbol{x}_\text{c}} - {\hat{\boldsymbol{x}}_{\text{AE}}}$. Eve constructs the test statistic by match-filtering the residual signal with the challenge signal $\boldsymbol{x}_\text{c}$  as
\begin{align}\nonumber 
	{\delta _{\text{TBCR,E}}} = \Re \{ \boldsymbol{x}_\text{c}^{\dag}{\boldsymbol{r}_\text{E}}\}.
	\label{TBCREDELTA}
	\tag{C.4}
\end{align}
Thus, the probability distribution of the test statistic at Eve when he is close to Alice can be modeled as 
\begin{align}\nonumber
	{\delta _{\text{TBCR,EA}|{\text{H}_0}}}\sim \mathcal{CN}\left(0,\frac{L\left(\sigma_\text{A}^2+\sigma_\text{E}^2\right)}{2|h{|^2}}\right),	
	\tag{C.5}
\end{align}
\vspace{-0.1cm}
and
\vspace{-0.1cm}
\begin{align}\nonumber
	{\delta _{\text{TBCR,EA}|{\text{H}_1}}}\sim \mathcal{CN}\left(\left(1-\rho_\text{s}\right)L,\frac{L\left(\rho_\text{s}^2\sigma_\text{A}^2+\sigma_\text{E}^2\right)}{2|h{|^2}}\right).
	\tag{C.6}
\end{align}
In high SNR regions, where $P_\text{D}\geq$ 50\% holds, the threshold $\gamma$ satisfies $\Gamma\!=\!\left(\gamma-\left(1-\rho_\text{s}\right)L\right)<0$. For sufficiently small $\sigma_{\text{A}}^2$, we have 
\begin{align}\nonumber
	Q\left(\frac{\Gamma\sqrt{2|h{|^2}}}{\sqrt{L\left(\rho_\text{s}^2\sigma_\text{A}^2+\sigma_\text{E}^2\right)}}\right)\!<\!	Q\left(\frac{\Gamma\sqrt{2|h{|^2}}}{\sqrt{L\sigma_\text{E}^2}}\right),\quad \sigma_\text{A}^2\!\rightarrow\! 0.
	\tag{C.7}
\end{align}
Thus, the asymptotic upper bound for Eve'performance in high SNR regions can be obtained by ignoring the noise of Alice.

Following the above method, when Eve approaches Bob, the received signal can be simplified as
\begin{align}\nonumber
	{\boldsymbol{y}_{\text{BE}|{\text{H}_0}}} = h\left(h^*{\boldsymbol{x}_\text{c}} + {\boldsymbol{w}_{\text{A}}}\right) + \boldsymbol{w}_{\text{E}},
	\tag{C.8}
\end{align}	
\vspace{-0.1cm}
and
\vspace{-0.1cm}
\begin{align}\nonumber
	{\boldsymbol{y}_{\text{BE}|{\text{H}_1}}} = h\left({\rho_\text{s}}\left(h^*{\boldsymbol{x}_\text{c}} + {\boldsymbol{w}_{\text{A}}}\right) + {\rho_\text{t}}\boldsymbol{t}\right) + {\boldsymbol{w}_{\text{E}}}.
	\tag{C.9}
\end{align}
Based on the cascaded channel $H=|h|^2$, Eve performs matched filtering on the received signal to obtain
\begin{align}\nonumber
	{\hat{\boldsymbol{x}}_{\text{BE}|{\text{H}_0}}} = {\boldsymbol{x}_\text{c}} +\frac{{\boldsymbol{w}_{\text{E}}}}{|h|^2},
	\tag{C.10}
\end{align}
\vspace{-0.1cm}
and
\vspace{-0.1cm}
\begin{align}\nonumber
	{\hat{\boldsymbol{x}}_{\text{BE}|{\text{H}_1}}} = {\rho_\text{s}}{\boldsymbol{x}_\text{c}} + \frac{{\rho_\text{t}}\boldsymbol{t}}{h}+ \frac{{\boldsymbol{w}_{\text{E}}}}{|h|^2}.
	\tag{C.11}
\end{align}
Then, Eve constructs the test statistic as in (\ref{TBCREDELTA}), where ${\boldsymbol{r}_\text{E}} = {\boldsymbol{x}_\text{c}} - {\hat{\boldsymbol{x}}_{\text{BE}}}$. 
The process of obtaining false alarm probability, optimal threshold and detection probability is similar to the derivation of (\ref{eqTBCR_PFA}) and (\ref{eqTBCR_PD}) from (\ref{delta_TBCR_H0}) and (\ref{delta_TBCR_H1}) in Appendix A, and will not be reiterated here.

The proof is completed.
 $\hfill\blacksquare$

\vspace*{-10pt}

\bibliography{ylbib.bib}
\bibliographystyle{IEEEtran}
\end{document}